\documentclass[12pt]{iopart}
\usepackage{txfonts}
\usepackage{bm}
\usepackage{graphicx}
\usepackage{iopams}
\usepackage{mathrsfs}
\usepackage[sort&compress, numbers, comma]{natbib}
\usepackage{feynmf}  
\newcommand{\SB}{\texttt{SuperBayeS}}
\newcommand{\eqref}[1]{(\ref{#1})}
\newcommand{\Cp}{{\mathscr{C}}}
\newcommand{\SMp}{{\mathscr{N}}}
\newcommand{\Einastop}{{\mathscr{H}}}
\newcommand{\Velp}{{\mathscr{V}}}
\newcommand{\be}{\begin{equation}}
\newcommand{\ee}{\end{equation}}
\newcommand{\like}{\mathcal{L}}
\newcommand{\Vmax}{V_{\rm max}}
\newcommand{\rmax}{r_{\rm max}}
\newcommand{\data}{d}

\begin{document}

\title[Indirect Dark Matter Detection from Dwarf Satellites]{
Indirect Dark Matter Detection from Dwarf Satellites: \\
Joint Expectations from Astrophysics and Supersymmetry}

\author{Gregory D. Martinez}
\address{Center for Cosmology, Department of Physics and Astronomy, 
University of California, Irvine, CA,  92697}
\ead{gmartine@uci.edu}

\author{James S. Bullock}
\address{Center for Cosmology, Department of Physics and Astronomy, 
University of California, Irvine, CA,  92697}
\ead{bullock@uci.edu}

\author{Manoj Kaplinghat}
\address{Center for Cosmology, Department of Physics and Astronomy, 
University of California, Irvine, CA,  92697}
\ead{mkapling@uci.edu}

\author{Louis E. Strigari}
\address{
Kavli Institute for Particle Astrophysics and Cosmology, Physics Department, 
Stanford University, Stanford, CA,  94305}
\ead{strigari@stanford.edu}

\author{Roberto Trotta} 
\address{ 
Imperial College London, Astrophysics Group, Blackett Laboratory, Prince Consort Road, London SW7 2AZ, UK}
\ead{r.trotta@imperial.ac.uk }

\begin{abstract}

We present a general methodology for determining the gamma-ray flux from annihilation
of dark matter particles in Milky Way satellite galaxies, focusing on two promising 
satellites as examples: Segue 1 and Draco. We use the SuperBayeS code to explore the 
best-fitting regions of the Constrained Minimal Supersymmetric Standard Model (CMSSM) 
parameter space, and an independent MCMC analysis of the dark matter halo properties 
of the satellites using published radial velocities. We present a formalism for 
determining the boost from halo substructure in these galaxies and show that its value
depends strongly on the extrapolation of the concentration-mass (c(M)) 
relation for CDM subhalos down to the minimum possible mass. 
We show that the preferred region for this minimum halo mass within the CMSSM with neutralino dark matter is 
$\sim 10^{-9} -10^{-6}$ M$_\odot$. For the boost model where the observed power-law 
c(M) relation is extrapolated down to the minimum halo mass we find average boosts 
of about 20, while the Bullock et al (2001) c(M) model results in boosts of order unity.
We estimate that for the power-law c(M) boost model and photon energies greater than a GeV, 
the Fermi space-telescope has about 20\% chance of detecting a dark matter annihilation 
signal from Draco with signal-to-noise greater than 3 after about 5 years of observation. 

\end{abstract}

\maketitle

\section{Introduction}
If the dark matter in the Universe 
is comprised of stable, weakly-interacting massive particles (WIMPs), in many instances
this leads to the prediction that WIMPs will self-annihilate into Standard Model (SM) particles
that may be visible with the upcoming generation of high-energy particle detectors. 
If high energy gamma-rays are produced, there are several promising 
sources within our own Galactic environment where the annihilation radiation 
from WIMPs may be visible. 
A detection of annihilation products from multiple sources, in possible concert with 
detections from colliders and underground labs, will be required to conclusively 
establish the nature of the dark matter in the Universe.

Each source of annihilation radiation
has its advantages and disadvantages. Because of its close proximity and high dark matter density, 
the flux from annihilation radiation will be the largest in the direction of the
Galactic center. 
However, uncertainties in the empirical determination of the central density profile
~\cite{Bergstrom98,Binney2001,Klypin2002,Widrow:2005bt} 
and in the contamination from gamma-ray 
sources that are not of dark matter origin~\cite{Jeltema:2008hf} 
may hinder the extraction of a dark matter signal 
from this region. Both of these systematics may be somewhat alleviated by searching 
for annihilation 
radiation at a few degrees offset from the Galactic center~\citep{Stoehr:2003hf}, but 
even in this region a full analysis and understanding 
of the spectrum of the astrophysical backgrounds is required. 

The high mass-to-light ratios and the relative proximity of the dwarf spheroidal 
satellite galaxies (dSphs) of the Milky Way make them also excellent independent targets that have been widely 
considered in the literature~\citep{Baltz00,Tyler:2002ux,Evans:2003sc,Strigari:2006rd,
BergstromHooper06,Pieri04,Sanchez-Conde07,Colafrancesco07,Bertone07,Pieri08}. Their status as potential targets for indirect detection
has become even more interesting recently, given that the known number of satellites has more than doubled
in the past few years
~\cite{Willman:2004kk,Willman:2005cd,Zucker:2006he,Zucker:2006bf,Belokurov:2006ph}, 
coupled with the discovery that all of these satellites share a common dark 
matter mass scale within their central 300 pc~\cite{Strigari:2008ib}. 
While the actual signal of gamma-ray 
flux from the Milky Way dSphs is smaller than the flux from the Galactic center, 
the astrophysical gamma-ray backgrounds tend to be reduced in the direction of 
these objects, as most of them are located at high Galactic latitudes. Additionally, these dSphs 
have low intrinsic emission from astrophysical gamma-ray sources; not only do they have
mass-to-light ratios greater than $\sim 100$ in most cases, but 
all of them within $\sim 400$ kpc have strong upper limits
on HI gas content~\cite{Grcevich:2009gt}. 

The astrophysical contribution to the calculation of the annihilation flux from any  
source can generally be divided into two components: one component arising from the smooth halo
contribution, which is proportional to the density squared distribution in the halo, and one component
arising from the flux due 
to bound substructures within each of these halos. While 
it has long been recognized that the presence of substructure 
in dark matter halos can have a significant effect on
the annihilation rate of dark matter particles~\citep{Silk:1992bh}, theoretical calculations of this boost factor from substructure have varied by orders of magnitude
because of large uncertainties in both the density profile of substructures and their distribution within the parent halos
~\cite{Baltz00,Strigari:2006rd,PieriBertone08,Afshordi:2008mx}. Numerical simulations are now reaching the necessary resolution to characterize
the density distribution of substructures~\cite{Kuhlen:2008aw,Springel:2008cc}, resolving substructure down to four
levels in the hierarchy~\citep{Springel:2008cc}. 
These, the highest resolution simulations, model the evolution of the dark matter and do not include baryons.  
In principle, simulations of this kind can be used to provide accurate estimates for the expected boost factor only 
in objects with negligible baryonic mass fractions (e.g., dwarf galaxies).  However,  in regions like the Galactic center 
where baryons play an important role in the dynamics, baryonic effects, such as encounters of substructures with 
stars in the Galactic disk or bulge
~\citep{Angus:2006vp}, or the backreaction of the dark matter distribution in response to disk formation
~\citep{Abadi:2009ve}, must be included.  Future simulations that include more physics and even higher 
resolution will be required to fully characterize boost factors in general circumstances.

It is clear that, in order to extract an unambiguous detection of dark matter annihilation radiation, 
a full understanding of all astrophysical uncertainties is required. 
With a goal of characterizing these uncertainties, in this paper we present 
an algorithm for the calculation of the gamma-ray flux 
from dwarf satellites, accounting for both uncertainties in the smooth halo distribution and the
halo substructure distribution. We introduce a method for scanning the parameter
space and determining the best fitting dark matter distributions from the kinematics 
of stars in these satellites using a Monte Carlo Markov Chain (MCMC) analysis.  
We combine the parameter constraints from the satellite stellar kinematics with the constraints
on the parameters of the Constrained Minimal Supersymmetric Standard Model (CMSSM),
with a goal of obtaining  predictions accounting in a realistic way for all relevant sources of uncertainty 
of the flux from the dwarf 
satellites. The predicted regions that we delineate will provide guidance for future 
gamma-rays experiments for testing the predictions of neutralino dark matter in the CMSSM
self-consistently within the context of CDM. As examples, we apply our algorithm 
to two particular satellites, both of which are known to be strongly
dark matter dominated: the classical satellite 
Draco, which is located at 80 kpc from the Sun, 
and Segue 1, which is a newly-discovered satellite
at 23 kpc from the Sun. We provide these two as examples and leave the work of ranking all
the satellites in terms of their expected gamma-ray luminosity to a future paper.

As a part of our analysis, we provide a new analytic solution to the
equation governing the boost factor from halo substructure. Our solution is 
particularly useful because it allows for mass functions and halo
concentrations to be free functions of host halo mass that can be implemented
at each level in the substructure hierarchy. This is particularly important 
when implementing results from recent numerical simulations which show 
that the normalization of the mass function is reduced by up to 50\% at the next level
of hierarchy~\cite{Springel:2008cc}. Using this solution, we show that the uncertainty in the
boost factor is dominated by the extrapolation of the dark matter halo concentration versus
mass relation down to mass scales that are currently unable to be resolved in CDM
simulations. Assuming an optimistic power law extrapolation, we find mean boost
factors $\sim 20$, in agreement with recent numerical extrapolations for Milky Way 
mass halos~\cite{Springel:2008cc}. 
Assuming a concentration-mass relation that is linked to the small-scale power spectrum
 as in the model~\citet{Bullock2001}, however,
leads to boost factors of order unity. 
As an additional application of this analytic formula, we solve for the minimum dark matter 
halo mass at each 
point in the CMSSM parameter space, and find that the typical range for the minimum mass
CDM halo is $10^{-9}-10^{-6}$ M$_\odot$. This result updates and generalizes the previous calculations
of the minimum mass halo in the context of Supersymmetric CDM
~\citep{Schmid:1998mx,Hofmann:2001bi,Loeb:2005pm,Profumo:2006bv,Bertschinger06}. 

The paper is organized as follows. In~\Sref{sec:annihilation} we 
review the formalism for determining the gamma-ray flux from
dark matter annihilations. In~\Sref{section:CMSSM},
we review our assumptions for the CMSSM and the \SB~
software that scans the CMSSM parameter space. 
In~\Sref{section:dwarf} we review the formalism to determine 
the best fitting dwarf satellite dark matter halo profiles, under the
assumption of the CDM model. In~\Sref{section:boost}
we present our calculation of the probability distribution for the
boost factor and the resulting differences relative to the smooth
halo flux. In~\Sref{section:results} we discuss some 
detection prospects for present observatories, and in~\Sref{section:conclude} 
we present our conclusions. Our main conclusions for the detectability of the
flux are summarized in Figure~\ref{fig:1Dfluxes}. 

\section{Annihilation Flux and Gamma-ray spectrum} 
\label{sec:annihilation} 
Following standard methods~\citep{Jungman96,Bergstrom98} 
the gamma-ray flux from particle annihilations can be derived via 
\begin{equation}
\label{eq:flux}
 \Phi(E) = \frac{<\sigma \nu> N_{\gamma}(E)}{8 \pi m_{\chi}^2}
 \int^{\theta' = \theta_{\rm max}}_{\theta'=0} d\Omega' \int d\Omega \mathcal{R}\left(\vec{\theta'}-\vec{\theta}\right) \int_{\ell_{-}}^{\ell_{+}} 
\rho_{DM}^2[\ell(\theta)]d\ell(\theta),
\end{equation}
where $\ell$ is the line-of-sight distance, $\ell_{\pm} = d
\cos \theta \pm \sqrt{r_{t, DM}^2 - d^2 \sin^2 \theta}$, $d$ is the
distance to the galaxy, $\theta$ is the line-of-sight angle from the center of
the galaxy, $<\sigma \nu>$ is the average cross section for
annihilation, $r_{t, DM}$ is the tidal radius of the dark matter
halo, and $m_{\chi}$ is the WIMP mass.  Here,
\begin{equation}
 N_{\gamma}(E) = \int_{E}^{m_{\chi}}\frac{d N}{d E}dE
\end{equation}
is the number of photons above energy $E$ produced per annihilation and
the resolution window function is
\begin{equation}
\mathcal{R}\left(\vec{\theta}\right) = \frac{\ln 2}{4 \pi \theta_{res}^2} \exp\left[-\ln2 \frac{\vec{\theta}^2}{ \theta_{res}^2}\right].
\end{equation}
For Fermi, $\theta_{res}$ is approximately 10 arcminutes at the energies we
consider. 
It is clear from~\Eref{eq:flux} that an accurate prediction for the flux entails incorporating relevant uncertainties both for the astrophysical quantities and for the particle physics model. In the present analysis we constrain the 
density of the dark matter halo, $\rho(\ell)$, from the kinematics of the stars, 
while the annihilation cross section, WIMP mass and annihilation channels 
we use are derived in the following section. 

\section{The Constrained Minimal Supersymmetric Standard Model}
\label{section:CMSSM}

For the calculations in this paper, we will assume that the dark matter 
particle consists of the lightest stable supersymmetric particle. More 
specifically, we focus on the case of the CMSSM. In this section, 
we review the basic definitions and parameters of the CMSSM, 
discussing specifically the \SB~ code that we use to explore the CMSSM
parameter space and how this feeds into our calculation of the flux.

\subsection{Neutralino Dark Matter in the CMSSM}

Supersymmetry (SUSY) provides a compelling and well-motivated extension to
the Standard Model that naturally contains a dark matter
candidate~\citep{Jungman96}. Supersymmetry postulates a symmetry
between bosons and fermions -- every boson (e.g. gauge bosons) has a
fermonic partner (e.g. gauginos) and every fermion (leptons, quarks)
has a bosonic partner (sleptons, squarks). The particularly well-studied R-parity
conserving, weak-scale softly broken SUSY models provide
both natural solutions to the ``fine tuning problem'' and a natural
dark matter candidate \citep{Martin98}.  The former is achieved
through cancellation of quadratic divergences of one loop quantum
corrections to the Higgs mass.  The latter is due to a conserved 
discrete symmetry (R-parity) that prohibits the lightest SUSY particle 
from decaying to only SM particles.

Enlarging the
particle sector in this manner greatly increases the number of free 
parameters that specify the model; 
even the most minimal form of SUSY (MSSM) introduces over a
hundred new parameters.  
Such a large number of free parameters  
makes the efficient exploration of the MSSM parameter
space very challenging. The naive method of exploring the likelihood surface on a regularly--spaced 
grid is clearly inadequate, as the required computational effort scales exponentially with the number 
of dimensions considered. Furthermore, present--day constraints on SUSY phenomenology are fairly indirect 
and do not allow for meaningful constraints on models with so many degrees of freedom. A popular and 
well motivated simplification is achieved by demanding SUSY parameter unification at GUT (Grand Unified Theory) scales.  
This so-called constrained MSSM (CMSSM) now limits the number of SUSY
parameters to 4 continuous and 1 discrete parameter \citep{Kane93}: the common gaugino mass, $m_{1/2}$,
common mass for scalars, $m_0$, trilinear scalar couplings, $A_{0}$, (all of which are specified at the GUT scale, $M_{\rm{GUT}} \simeq 2\times 10^{16}$GeV) the 
ratio of the higgs expection values, $\tan\beta$, and the sign of the
``$\mu$ term'', sgn($\mu$)~\citep{Martin98, Kane93}. We shall denote the CMSSM parameters as 
\begin{equation}
\Cp \equiv \{  m_0, m_{1/2}, A_0, \tan\beta, \rm{sgn}(\mu) \} \, .
\end{equation}
It has been recently demonstrated~\cite{Allanach:2005kz,Austri06,Roszkowski:2007fd,Feroz:2008wr,Trotta08} that 
the value of some SM parameters can strongly affect the predictions for some of the observable quantities, 
in particular the relic neutralino abundance (see Fig.~4 in~\citet{Roszkowski:2007fd}). Therefore, it is not 
sufficient to specify the values of $\Cp$ and fix the relevant SM parameters to their current best--fit values, but the 
latter must be introduced as ``nuisance parameters'' and marginalized over, in order to account for their impact in the 
predictions. The most relevant SM parameters are 
\begin{equation}
\SMp \equiv \{  m_t, m_b, \alpha_S, \alpha_{em} \} \,, 
\end{equation}
namely, the top quark mass, the bottom quark mass, the
strong coupling constant and electromagnetic coupling constant, respectively.

In the context of the CMSSM, specification of the parameter set $(\Cp, \SMp)$ allows for derivation of a 
full suite of predictions for low--energy observables. The package~\SB~, developed by
~\citet{Austri06} and~\citet{Trotta08} embeds several codes in a MCMC framework to derive from 
$(\Cp, \SMp)$  the SUSY mass spectrum (via SoftSusy \citep{Allanach02}), the neutralino relic abundance 
(via~DarkSusy~\citep{Gondolo04} or MicrOMEGAs~\cite{Belanger:2001fz,Belanger:2004yn}), SUSY 
corrections to various Higgs sector quantities (employing FeynHiggs \citep{Frank07, Degrassi03,
Heinemeyer99, Heinemeyer00}) and branching ratios of rare processes (using Bdecay~\citep{Foster05}). 
The CMSSM and SM parameter space is explored by \SB~using an MCMC Metropolis--Hastings algorithm, 
or, more recently, by employing the more efficient and robust ``nested sampling'' 
algorithm~\cite{Skilling04, Feroz:2007kg, Trotta08}. The parameters are then constrained by applying all 
available constraints on the low--energy observables, including the WMAP 5-years determination of the 
relic abundance, sparticle and Higgs mass limits, branching ratios of rare processes, electroweak observables 
and direct constraints on the SM quantities (for a detailed discussion of the likelihood, 
see~\cite{Austri06,Roszkowski:2007fd,Trotta08}). 

\subsection{Priors in the CMSSM}

The final outcome of the CMSSM analysis is a list of samples drawn from the posterior distribution, $P(\Cp, \SMp | D)$ obtained via Bayes' theorem: 
\be \label{eq:bayes}
P(\Cp, \SMp | D) \propto \like(\Cp, \SMp) P(\Cp, \SMp) \, ,
\ee
where $D$ denotes the combined data described above, $\like(\Cp, \SMp) \equiv P(D | \Cp, \SMp) $ is the likelihood function and $P(\Cp, \SMp)$ is the prior distribution for the CMSSM and SM parameters. From the posterior one can then derive the probability distribution for any function of the quantities $(\Cp, \SMp)$ one is interested in, for example the neutralino--proton interaction cross--section (relevant for direct dark matter detection experiments, see~\cite{Trotta07}), the gamma-ray and antimatter flux from the galactic center (relevant for indirect detection searches~\cite{Roszkowski:2007va}) and Higgs--sector physics~(of interest for Higgs-boson searches~\cite{Roszkowski:2006mi}).

The posterior in~\Eref{eq:bayes} should be dominated by the likelihood, $\like$, so that the prior influences vanish  
for strongly constraining data (for more details, see e.g.~\cite{Trotta:2008qt}). However, it has been found that this 
is not currently the case for the CMSSM --- namely, the available data are not sufficiently constraining to 
determine the CMSSM parameters in a prior--independent way (see~\citet{Trotta08} for a detailed analysis). This 
means that some of the posterior constraints on $\Cp$ are somewhat dependent on the chosen prior distribution. 
The fundamental reason for this is that the mapping between high--energy CMSSM parameters and low energy 
observable quantities is highly non--linear, due to the nature of the Renormalization Group Equations, and therefore 
even fairly strong low-energy constraints are relatively mildly informative about the quantities one is interested in, 
namely $\Cp$. It is however expected that this issue will be resolved once the LHC will deliver direct observations 
of the SUSY mass spectrum~\cite{rrt4}.  

\subsection{CMSSM samples and derived quantities}

For the results in this paper, we use a nested sampling chain for the parameter space spanned by $(\Cp, \SMp$), containing 
approximately $45,000$ samples. We assume throughout a positive sgn($\mu$)  (motivated by consistency with 
the measured anomalous magnetic moment of the muon). We adopt the chain resulting from the analysis 
in~\citet{Trotta08}, with a flat prior with the limits denoted in~\tref{tab:limits} (the ``4 TeV'' limits in~\citet{Austri06}). 

From those CMSSM chains, we derive for each sample the value of the WIMP mass, $m_\chi$, its annihilation 
cross section,  $<\sigma \nu>$, and number of photons 
produced in the annihilation above 1 GeV, $N_{\gamma}(1$
GeV$)$.  As discussed below, we choose
1 GeV because it provides a conservative lower bound for the 
expected signal energy window from CMSSM dark matter annihilation. Furthermore, we 
 compute the value of the minimum mass halo, $m_{\rm min}$, as explained in section~\ref{sec:mmin} below. 

\section{Dwarf Satellite Kinematics\label{section:dwarf}}

We follow the standard formalism for analyzing stellar line-of-sight
velocities from dwarf satellites. In this section, we
review the relevant formulae so as to establish notation and
conventions that we use throughout this paper. 

\subsection{Theoretical Modeling} 

We consider the satellites as two-component systems
consisting of stars and dark matter
(e.g.~\cite{Lokas:2004sw,Strigari07-redef}). 
The potential is assumed to be spherically symmetric, and
the system is taken to be completely pressure-supported (no rotational
support). This is seen to be a good 
description of many of the dwarf satellites
~\cite{Walker2006,Koch:2006in,Koch:2007ye}. 
With these assumptions the Jeans equation is
\begin{equation}
\label{eq:jeans}
r\frac{d(\rho_{\star} \sigma_r^2)}{d r}  = -\rho_{\star}\frac{G M(r)}{r}  - 2
\beta \rho_{\star} \sigma_r^2,
\end{equation}
where $\rho_{\star}$ is the stellar density, $\sigma_r$ is the stellar
radial velocity dispersion,
$\beta \equiv 1 - \sigma_t^2/\sigma_r^2$ is the velocity anisotropy
parameter,  and $\sigma_t$ is the stellar tangential velocity
dispersion. The mass of the system, $M(r)$, is defined as the
total dynamical mass, which is the sum of the contributions 
from the stars and the dark matter. The 
line-of-sight velocity dispersion is 
\begin{equation} \label{eq:sigmalos}
\sigma_{th} = \frac{2}{I_{\star}(R)}\int_R^{\infty}\left(1 -
\beta\frac{R^2}{r^2}\right)\frac{r \rho_{\star}\sigma_r^2}{\sqrt{r^2 -
R^2}} dr
\label{eq:LOS}
\end{equation}
where $R$ is the projected radius onto the sky and $I_\star(R)$ is the 
stellar surface density. The use of subscript ``{\it th}" will become apparent
below when~\Eref{eq:LOS} is fed into our statistical analysis. 

Observationally, the integrated  
masses within $\sim 300$ pc for all of the Milky Way 
satellites are very similar, consistent with 
$\sim 10^7$ M$_\odot$ independent of the
dwarf galaxy luminosity~\cite{Strigari:2008ib}. To first order, this fact simplifies
the selection of the best targets for flux detection to those that are
closest to the Sun~\cite{Strigari:2007at}. However,
as we discuss below, including the effects of a prior relation for the maximum 
circular velocity distribution for dark matter halos somewhat modifies
this simple estimate.  In general, the primary factors that determine
the best sources are those objects that 
have the best signal-to-noise ratio, accounting for the astrophysical
backgrounds. Some often-discussed targets include Segue 1
(23 kpc), Ursa Major II (32 kpc), Willman 1 (38 kpc), and
Coma Berenices (44 kpc) (e.g.~\cite{Strigari:2007at,Bringmann:2008kj,Geha2008})
all of which were discovered since 2004. 
The half-light radii of these objects are $\sim$ 10-100 pc, and given 
their high velocity dispersions of $\sim 4-6$ km/s~\cite{SimonGeha2007},
these objects are consistent with 
being dark-matter dominated dSphs~\cite{SimonGeha2007,Strigari:2008ib,Geha2008}. 
The nearest of the more well-known (classical) dSphs include Ursa Minor (66 kpc) and Draco 
(80 kpc); previous calculations show that these two objects have similar predicted fluxes~\cite{Strigari:2006rd}. 

Since our main goal in this paper is to discuss the methodology 
for robustly predicting fluxes and 
including the boost in the calculation of the gamma-ray flux, 
we restrict our analysis to two example dSphs: 
Draco and Segue 1. The kinematic constraints we derive for Draco are much 
stronger, as this object has been well-studied both from the standpoint
of its photometry and kinematics. Segue 1 is a more-recently discovered
satellite that appears to be strongly dominated by dark matter and the least 
luminous galaxy known~\cite{Geha2008}. However, there are only 24 stars with measured
velocities in Segue 1~\cite{Geha2008}, and, as we show below, the errors on its mass and flux are much
larger than the respective values for Draco. The surface densities of these objects 
are fit by King~\citep{King62}
and Plummer~\citep{Plummer11} profiles, respectively. For Draco, the King
core radius is $r_{king} = 0.18$ kpc, and the 
King tidal radius is $r_t = 0.93$ kpc~\citep{Odenkirchen:2001pf}. 
For Segue 1, the one-component Plummer fit gives a Plummer
core radius of $r_{pl} = 35$ pc~\cite{Martin2008}. 
In the Plummer profile, $\rho_\star$ falls off
as $1/r^5$ in the outer regions, so there is 
no natural definition of the stellar tidal radius, 
in contrast with the King profile. In this case, 
we conservatively assume that the stellar tidal radius
is given by the position of the outermost observed
star, which is located at a projected radius of $R = 50$ pc~\cite{Geha2008}. 
In principle, the stellar surface density parameters could also be estimated jointly with our other 
model parameters, however in the present work we choose to fix them instead to their best fit values as given above. 

To model the density profile of the respective dark matter halos
of these objects,  we use an Einasto profile, which is defined as 
\begin{equation}
\rho(r) = \rho_s \exp 
\left\{-2n\left[\left(\frac{r}{r_s}\right)^{1/n}-1\right]\right\}. 
\label{eq:einasto}
\end{equation}
The Einasto profile 
has been shown to be a good fit to CDM halos with the Einasto index,
$n$, ranging from $\sim 3 - 7$ \citep{Merritt06-emp.model.1,
Navarro04-cdm.halo.3, Gao07}.  For our purposes, this profile is
also convenient for two separate reasons: (1) the profile has a well-defined
mass, which is important when we calculate substructure boost factors
below, and (2) the profile does not diverge towards the center of the
halo, which is convenient when calculating the gamma-ray
flux. As presented in~\Eref{eq:einasto}, 
there are three parameters in the Einasto profile that we must determine
from the data: the log-slope index $n$, the scale radius $r_s$, and the
scale density, $\rho_s$.  It will be convenient in the following to replace the Einasto density and radius variables ($\rho_s$, $r_s$) with 
the implied halo maximum circular velocity and radius of maximum circular velocity ($\Vmax$, $\rmax$).
 The dark matter halo density profile is then specified in terms of the parameter set
\begin{equation}
\Einastop \equiv \{  n, \Vmax, \rmax \} \, .
\end{equation}

The final quantity we must specify is the velocity 
anisotropy, which enters both directly in the Jeans equation
in~\Eref{eq:jeans} and in the equation that relates the observed line-of-sight velocity dispersion
to the underlying stellar radial velocity dispersion~\Eref{eq:LOS}. This quantity is unconstrained by line-of-sight velocity
data~\cite{StrigariBullockKaplinghat07}, so in order to allow for general models, 
we model the velocity anisotropy as
\begin{equation}
\beta(r) = \frac{\beta_0 + \beta_{\infty} 
\left(r/r_{\beta}\right)^{\eta}}{1 + \left(r/r_{\beta}\right)^{\eta}},
\label{eq:beta}
\end{equation}
with four free parameters, 
\begin{equation}
\Velp \equiv \{ \beta_0, \beta_\infty, r_\beta, \eta\} \, .
\end{equation}
We note that this parametrization is slightly more general than that used in
\citet{Strigari07-redef}, allowing here for the power law index $\eta$
in addition to the anisotropy scale radius, $r_\beta$, and the asymptotic
inner and outer anisotropies, $\beta_0$ and $\beta_\infty$. 
It should also be noted that there is an intrinsic 
degeneracy between the logarithmic slope of the density profile 
in~\Eref{eq:einasto} and anisotropy.  Even in the simplified case of 
constant anisotropy, this degeneracy restricts how well 
the central slope $n$ of the halo may be contrained from stellar kinematics~\cite{StrigariBullockKaplinghat07}.  

\subsection{Likelihood function of dSph model parameters}   

Above we have discussed our theoretical modeling; 
we now turn to our description of the line-of-sight velocity data~\cite{Walker2007,Geha2008}. 
We begin by noting that 
the line-of-sight velocities from dSphs are well-described by 
Gaussian distributions~\citep{Walker2006}. 
The observed velocity distribution is a 
convolution of the intrinsic velocity distribution, arising from the
distribution function, and the measurement uncertainty from an 
individual star. 
The probability of obtaining a set of data $\data$ given a set of model 
parameters ${\Einastop, \Velp}$ is described by the likelihood 
~\citep{Strigari:2006rd}
\begin{equation}
\label{eq:fulllike}
\like(\Einastop, \Velp) \equiv P(d| \Einastop, \Velp) =  \prod_{i=1}^{n} 
\frac{1}{\sqrt{2\pi(\sigma_{th, i}^2 + \sigma_{m, i}^2)}} 
\exp \left[-\frac{1}{2}\frac{(\data_i -u)^2}{\sigma_{th, i}^2 
+ \sigma_{m, i}^2}\right]\,, 
\end{equation}
where $\Einastop$ are the parameters that describe the 
density profile of dark matter and $\Velp$ are the stellar 
velocity anisotropy parameters.
The product is over the set of $n$ stars, and $u$ is the bulk
velocity of the galaxy in the direction of 
the observer.  As expected, the total error at a projected 
position is a sum 
in quadrature of the theoretical intrinsic dispersion, $\sigma_{th, i}(\Einastop, \Velp)$,
and the measurement error $\sigma_{m, i}$. 

Often, kinematic data from dSphs is presented in terms
of the velocity dispersion in bins of projected radius. In these
cases, it is useful to have an expression for the likelihood function
similar to~\Eref{eq:fulllike} that is free of any terms associated
with the measured velocities of individual stars. An expression of this form 
can be found by replacing $u$ and $\sigma$ by their respective maximum 
likelihood values, $\hat{u}$ and $\hat{\sigma}$, where the latter quantities
are obtained from a standard maximum likelihood procedure using
~\Eref{eq:fulllike}. 
Proceeding with this approximation, 
and also neglecting the measurement uncertainty in comparison with the
intrinsic dispersion (which is a good approximation for the bright
satellites), the likelihood function in 
~\Eref{eq:fulllike} can be reduced to
\begin{equation}
\label{eq:approxlike}
\like(\Einastop, \Velp) = \prod_{i=1}^{N} 
\frac{1}{\sqrt{2\pi} \sigma_{th, i}^{N_b}} \exp \left[-\frac{1}{2}
\frac{N_b \hat{\sigma}_b^2}{\sigma_{th, i}^2}\right].
\end{equation}
This is now a product over the number of bins in projected radius, $N$, 
each with a velocity dispersion $\hat{\sigma}_b$. The number of stars in 
a given bin is $N_b$.  
(Note the difference in the normalization relative to the expression presented
in~\citet{Strigari:2007at}, 
due to a typographical error in~\citet{Strigari:2007at}).
Note that the quantity $\sigma_{th, i} = \sigma_{th, i}(\Einastop, \Velp)$ is computed via~\Eref{eq:sigmalos}. 
As we did above in the context of the CMSSM parameters, we now go over to the posterior probability distribution 
function (pdf) for the parameters of interest, which is again given by Bayes' theorem 
\begin{equation}
\label{eq:posterior}
P (\Einastop, \Velp | d) 
\propto P(\Einastop)P(\Velp) \like(\Einastop, \Velp)
\end{equation}
where $P(\Einastop), P(\Velp)$ are the prior pdf's for the halo and velocity anisotropy parameters, respectively, which we take here to be uncorrelated. 
We will deal with the issue of priors in more detail below. 

The task is now to explore numerically the posterior,~\Eref{eq:posterior}, in order to determine credible regions 
on the model parameters. In previous similar work 
involving parameter estimation from dwarf satellites
~\citep{Strigari07-redef,Strigari:2007at}, the likelihood
function was directly integrated over the range of model
parameter space.
This method was accurate but could 
be time-consuming, particularly in the case of large
numbers of parameters to marginalize over. Rather than
direct numerical integrations, in this paper we 
explore the posterior distributions using Markov Chain Monte Carlo
techniques. Before discussing our MCMC methodology, we now 
turn to the discussion of the priors entering~\Eref{eq:posterior}.

\subsection{Priors for dSph parameters }
\label{subsec:priors}
\begin{figure}
\begin{center}
\rotatebox{270}{\includegraphics[height=0.45\hsize]{pics/mass30.ps}}
\rotatebox{270}{\includegraphics[height=0.45\hsize]{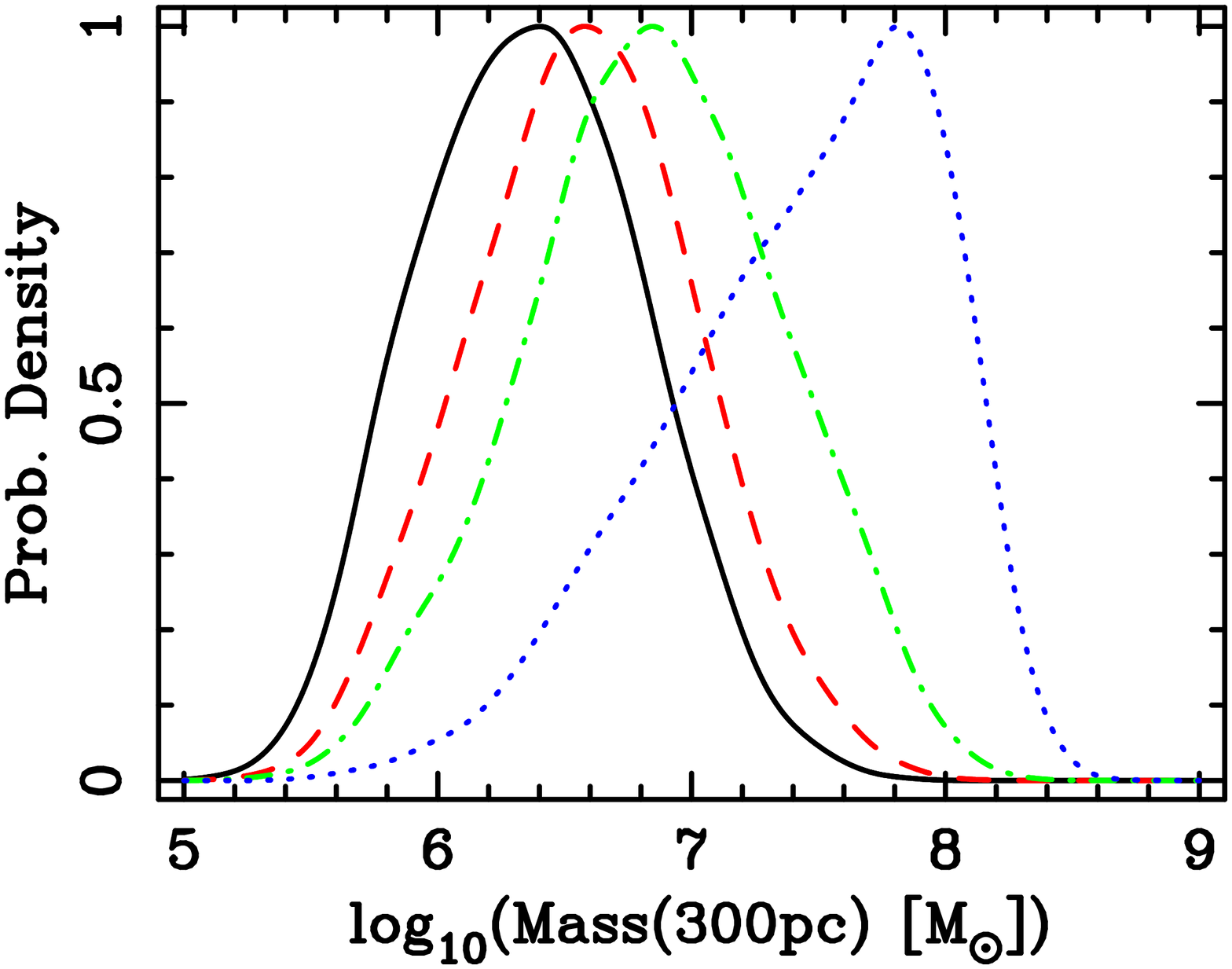}}
\caption{\footnotesize
Posterior probability distribution for the mass within 30 pc for Segue 1({\it left panel}) 
and the mass within 300 pc for Segue 1 ({\it right panel}). 
The four curves in each panel assume different Bayesian priors: 
a uniform prior in $V_{\rm max}^{-3}$ (black, solid)
$V_{\rm max}^{-2}$ (red, dashed), $V_{\rm max}^{-1}$ (green, dot-dashed), 
and $\ln(V_{\rm max})$ (blue, dotted). The prior distributions are truncated at 
$V_{\rm max} = 3 $km s$^{-1}$
as described in the text. Increasing negative powers of 
$V_{\rm max}$ causes the posterior to be more ``biased'' toward lower mass
solutions.  As a result, the posterior corresponding to these different priors differ.
\label{fig:prioreffect_seg}
}
\end{center} 
\end{figure}

\begin{figure}
\begin{center}
\rotatebox{270}{\includegraphics[height=0.45\hsize]{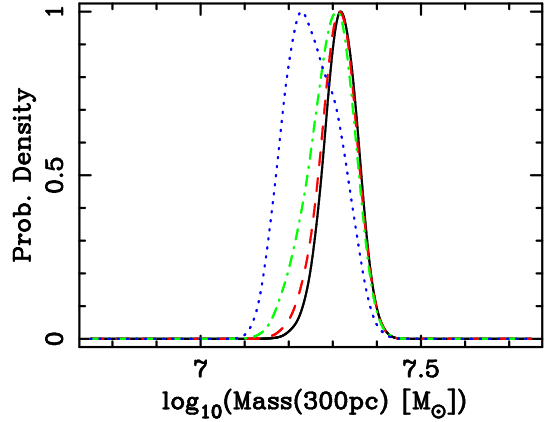}}
\caption{\footnotesize
Posterior probability distribution for the mass within 300 pc for Draco. 
The four curves in each panel assume different Bayesian priors: 
a uniform prior in $V_{\rm max}^{-3}$ (black, solid)
$V_{\rm max}^{-2}$ (red, dashed), $V_{\rm max}^{-1}$ (green, dot-dashed), 
and $\ln(V_{\rm max})$ (blue, dotted). The prior distributions are truncated at 
$V_{\rm max} = 3 $km s$^{-1}$
as described in the text. 
Note that the trend in mass within 300 pc with prior reverses here
compared to the case in Figure~\ref{fig:prioreffect_seg}.  This traces back in
part to the fact that the mass is best constrained near twice the half-light
radius~\cite{SKB07}. For Draco, 300 pc is within the half-light radius and
when the CDM $r_{\rm max}$-- $V_{\rm max}$ prior is imposed, lower
$V_{\rm max}$ halos are forced, on average, to be more concentrated at 300
pc. In Figure~\ref{fig:prioreffect_seg}, 300 pc is beyond the half-light radius
of Segue 1, and priors that favor larger $V_{\rm max}$ give larger
extrapolated masses.  
\label{fig:prioreffect_dra}
}
\end{center} 
\end{figure}

Choosing a prior in accordance with the physical situation and our degree of prior beliefs on it is an important aspect of Bayesian
analysis, as variations in the priors can lead to sizable differences in the posterior whenever the likelihood is not very strongly peaked. 
We account for this prior information in our notation by the inclusion of the appropriate
infinitesimals.  For example, a ``uniform'' prior probability in $x$
will be denoted as $P(x) = d(x)$ whereas a 
uniform prior probability in $\ln(x)$ is represented as 
$P(x) = d[\ln(x)] = d(x)/x$.
Regardless of these definitions, if data strongly constrains
some parameter, or a given combination of parameters, then the prior information
should not have much bearing on the result, as the posterior is dominated by the likelihood.
However as we will see for Segue 1, with only 24 radial velocities spanning only $\sim 50$ pc in radius, priors can have a significant effect. 

Guidance as to how to choose the priors for our dSph  model parameters can be gleaned from cold dark matter 
simulations which give precise predictions of halo abundances
given a halo shape and mass.  We note however, that these simulations
do not provide the probability of observing a {\em galaxy}, which itself depends 
not only on dark matter physics,
but also on star formation and baryonic physics.  But if we assume that primarily the halo mass, and not shape, affects
stellar physics we may draw the halo profile prior probabilities directly from simulations, 
with some additional simple inputs to account for gas physics. 

As discussed above, we describe the halo in terms of the parameters  $n, r_{\rm max}, V_{\rm max}$. 
We model the $r_{\rm max}$ prior (conditional on the value of $\Vmax$) as a log-normal distribution, which  
provides a good description of the $r_{\rm max}-V_{\rm max}$ relation measured in 
the Via Lactea \citep{Diemand:2008in} and Aquarius~\citep{Springel:2008cc} 
simulations over the entire range of $V_{\rm max}$. 
From the Aquarius simulations \citep{Springel:2008cc}, we estimate this relation, given
a $V_{\rm max}$, to be
\begin{equation}
\label{eq:conditionalprior}
 P(r_{\rm max} | V_{\rm max}) \propto \exp\left\{-\frac{\left[\log(r_{\rm max}) 
- 1.35\log(V_{\rm max})+1.75\right]^2}{2 \sigma_{\log(r_{\rm max})}^2}\right\} d (\log r_{\rm max}),
\label{eq:CDMprior}
\end{equation}
where $\sigma_{\log(r_{\rm max})} = 0.22$ is a conservative scatter
in the Aquarius subhalos for the entire range of 
$V_{\rm max}$.  There are no statistics published (as of this paper) for the parameter $n$ in the Einasto profile,
though from \citet{Springel:2008cc} it is
reasonable to assume a uniform prior in $1/n$ [i.e. $P(n) \propto d(1/n)$]
limited to the range $0.1 < 1/n < 0.5$.  

The choice of prior for $V_{\rm max}$ is a more delicate issue, for a couple of reasons.
One issue relates to the probability that a given CDM halo has a particular value of
$V_{\rm max}$, and a second issue relates to how well the line-of-sight velocity data itself 
constrain particular values of $V_{\rm max}$. On the former point,
$V_{\rm max}$ is the primary parameter that relates to the astrophysical
process of low-mass galaxy formation: small galaxies with more shallow potential wells are expected
to have low star formation rates, so the actual $V_{\rm max}$ prior is expected to
be more shallow than that inferred from the CDM subhalo mass function, $P(V_{\rm max}) \propto V_{\rm max}^{-4} dV_{\rm max}$ 
(derived from the relation $N(>V_{\rm max}) \propto V_{\rm max}^{-3}$ of \citet{Springel:2008cc}).
On the latter point, from the perspective of the line-of-sight data, large $V_{\rm max}$ values are not constrained 
by the data.  Given the form of the 
$r_{\rm max}$ prior, large $V_{\rm max}$ values correspond to large $r_{\rm max}$ values
that may fall outside the extent of the stellar profile.  Therefore, these values cannot be
directly constrained by data and accordingly become dominated by the prior.

\Fref{fig:prioreffect_seg} and~\Fref{fig:prioreffect_dra} exemplifies this behavior.  The left panel
of \Fref{fig:prioreffect_seg} shows 
the posterior constraints on the mass of Segue 1 within $30$ pc with four different
prior assumptions: $P(V_{\rm max}) \propto d(V_{\rm max}^{-3})$, $d(V_{\rm max}^{-2})$, 
$d(V_{\rm max}^{-1})$, and $d(\ln V_{\rm max})$.  The right panel shows the mass within $300$ pc using the
same respective priors.  Prior choice has little effect on $M(30$ pc$)$ since this value is
well constrained by line-of-sight velocity data.  In contrast the $M(300$ pc$)$ posterior  
is dominated by prior choice, as the radius of $300$ pc lies outside the measured stellar distribution.
In contrast Draco, whose stellar extent extends beyond $300$ pc, $M(300$ pc$)$ is well constrained.
We note that the prior behavior in \Fref{fig:prioreffect_seg} and~\Fref{fig:prioreffect_dra} does not contradict
the results presented in~\citet{Strigari:2008ib}, where a uniform prior in $\ln [M(300 {\rm pc})]$ was taken
for the entire satellite population, and no CDM-motivated priors were considered. 
Rather \Fref{fig:prioreffect_seg} and~\Fref{fig:prioreffect_dra} 
pertain to the specific case of the prior assumed in $V_{\rm max}$. 

The above situation does, however, present a dilemma when considering priors in the quantity $V_{\rm max}$:  
the actual V$_{\rm max}$ prior for the observable galaxy is likely to be more shallow
than the prior that comes from the predicted CDM $V_{\rm max}$ prior for all substructure (uniform in $V_{\rm max}^{-3}$), 
but more shallow priors will 
give more statistical weight to parts of $V_{\rm max}$
parameter space not well constrained by data.  Because of these factors, the best that can
be achieved with line-of-sight data is a lower limit to the model parameters.
Thus, in this paper we
use the CDM prior $P(V_{\rm max}) \propto V_{\rm max}^{-4} dV_{\rm max}$ 
 with an imposed cut-off of 3 km s$^{-1}$. This cut-off seems reasonable, as $\sim 3$
km s$^{-1}$ is expected to be a conservative lower bound to the 
$V_{\rm max}$ values below which gas is able to condense into halos~\cite{Thoul:1994ir}. 
As we show, the imposed cut-off of 3 km s$^{-1}$
does not affect the predicted flux from Draco, in which the posterior is dominated by the 
line-of-sight data. However, for Segue 1 (with only 24 stars), the issue is more subtle, as 
the CDM prior becomes more dominating with decreasing values of $V_{\rm max}$.  This is would be particularly 
true if the $V_{\rm max}$ cut-off extended down to arbitrarily low values, though our
choice of a (physically-motivated) cut-off somewhat reduces the effect of the prior, even for
Segue 1. 
In summary, our prior for the CDM halo parameters is given by
\be
P(\Einastop) = P(\rmax|\Vmax)P(\Vmax)P(1/\eta)
\ee 
with $P(\rmax|\Vmax)$ as in~\Eref{eq:conditionalprior} and $P(\Vmax), P(1/\eta)$ 
according to the prescriptions given above.

\begin{figure}
\begin{center}
\rotatebox{270}{\includegraphics[height=0.45\hsize]{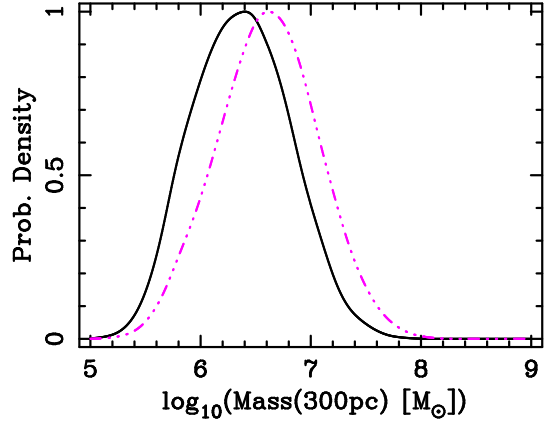}}
\caption{\footnotesize
Posterior probability distribution for the mass within 300 pc for Segue 1,
assuming an isotropic velocity distribution (the magenta dot-dot-dot-dashed line)
and the model in~\Eref{eq:beta} (black solid line). Both curves assume 
a uniform prior in $V_{\rm max}^{-3}$. 
Both curves assume a uniform prior in $V_{\rm max}^{-3}$ 
with an imposed cutoff below $V_{\rm max} = 3$~km s$^{-1}$.
\label{fig:prioreffect_beta}
}
\end{center} 
\end{figure}

Finally, we have little physical basis to
choose the prior for the anisotropy parameters $\Velp$.  Additionally, these
parameters are not well constrained by the data.
Fortunately, these parameters, unlike the halo parameters, do not have a direct
effect on the derived flux or mass (only indirectly through the Jeans equation). 
Thus, we expect that the choice of these priors will not
have as large of an impact on the result as compared to the halo priors.
\Fref{fig:prioreffect_beta} shows the difference between 
the Segue 1 mass withing $300$ pc using our prior assumption and the mass
assuming an isotropic velocity distribution.  Although our prior assumption
biases the probability distribution to lower mass values, the effect is not
as extreme as with the V$_{\rm max}$ prior.

The prior choices for all of the parameters (including particle physics quantities) considered in this paper are shown in~Table~\ref{tab:limits}.  

\begin{table}
\caption{\label{tab:limits}Summary of model parameters and the priors imposed on them. Unless otherwise stated, the prior pdf is flat within the given range.}
\begin{indented}
\item[]\begin{tabular}{@{}ccl}
\br
CMSSM Parameters, $\Cp$ & Priors Assumed & Notes\\
\mr
$m_0$ & $50$ GeV $ < m_0 < 4$ TeV & CMSSM scalar mass\\
$m_{1/2}$ & $50$ GeV $< m_{1/2} < 4$ TeV  & CMSSM gaugino mass\\
$A_0$ & $\vert A_0\vert < 7$ TeV  & Common Trilinear Coupling\\
$\tan\beta$ &$2 < \tan\beta < 62$  & Ratio of Higgs vev\\
\br
\br
SM Parameters, $\SMp$ & Priors Assumed & Notes\\
\mr
$M_t$ & $160$ GeV $ < M_t < 190$ GeV & Top quark mass\\
$m_b$ & $4$ GeV $ < m_b < 5$ GeV & Bottom quark mass\\
$\alpha_{em}$ & $127.5 < 1/\alpha_{em} < 128.5$ & EM coupling const.\\
$\alpha_s$ & $0.10 < \alpha_s < 0.13$ & Strong coupling const.\\
\br
\br
DM Halo Parameters, $\Einastop$ & Priors Assumed & Notes\\
\mr
$n$ & $0.1 < 1/n < 0.5$ & Einasto index, see~\Eref{eq:einasto}\\
$r_{\rm max}$ & see~\Eref{eq:CDMprior} & Used to derive $r_{S}$ \\
$V_{\rm max}$ & flat in $V_{\rm max}^{-3}$; $V_{\rm max} > 3$km/s & Used to derive $\rho_{S}$ \\
\br
\br
Anisotropy Parameters, $\Velp$ & Priors Assumed & Notes\\
\mr
$\eta$ & $0 < \eta < 3$ & see~\Eref{eq:beta}\\
$r_{\beta}$ & $0.01$ kpc $< r_{\beta} < 100$ kpc  & Anisotropy scale length\\
$\beta_0$ & $-2 < \beta_0 < 0$ & Central anisotropy\\
$\beta_{\infty}$ & $-3 < \beta_{\infty} < 1$ & Outer anisotropy\\
\br
\end{tabular}
\end{indented}
\end{table}

\subsection{Monte Carlo Markov Chain methodology}

Here we review the MCMC formalism necessary for our analysis;
we refer to the papers referenced for more details. 

The goal of an MCMC algorithm is to generate a series of points in parameter space (called ``a chain'') with the property that their density is distributed according to the posterior pdf one wishes to explore.  
Then from the chain of ``accepted'' points, the marginal probability 
distribution for each of the parameters is recovered by 
simply binning the points in the chain, and ignoring the uninteresting coordinates (two-dimensional distributions
are obtained in a similar manner). In our case, we wish to explore the joint parameter space spanned by the particle physics model parameters, $\Cp$ and $\SMp$, and by the dwarfs model parameters, $\Einastop$ and $\Velp$. So we are dealing with a total of 15 parameters. Because the CMSSM likelihood and the stellar kinematics likelihood are independent, the joint posterior factorizes as (more details are given below) 
\be \label{eq:jointP}
P(\Cp, \SMp, \Einastop, \Velp|D,d) \propto P(\Cp, \SMp)P(\Einastop, \Velp) \like(\Cp, \SMp)\like(\Einastop, \Velp) \, .
\ee
A great advantage of the MCMC procedure lies in it efficiency, for the computational effort scales roughly proportionally with the dimensionality of the parameter space being explored, rather than exponentially. The true
power of MCMC methods, which we specifically utilize in this paper,  
lies in the fact that in addition to obtaining the distributions for the model
parameters, the probability distribution for any
function of the model parameters is obtained by simply 
determining the function at each of the accepted points in the chain. 
In this way, we may easily determine the distribution of any 
parameter that is derived from our base set of parameters
by post-processing the chain of accepted points.
Examples of derived parameters that we are interested in for the purposes
of this paper include the dark matter mass of the dwarfs
or the gamma-ray flux. While the former quantity is just a function
of the parameters that describe the dark matter halo, $(\Einastop, \Velp)$
the latter quantity requires us to combine the 
probability distributions on $(\Cp, \SMp)$  with the probability
distribution from the dwarf kinematics. 
To understand how these probabilities can be combined, and quantities such as
the flux can be robustly calculated, we appeal to  
the general properties of conditional distribution functions. 

A Markov chain of the joint posterior distribution $P(\Cp, \SMp, \Einastop, \Velp|D,d)$,~\Eref{eq:jointP}, can be obtained from
the combination of the two posteriors  $P(\Cp, \SMp|D)$ and  $P(\Einastop, \Velp|d)$ as long as the two joint probability 
distribution can be factorized as in~\Eref{eq:jointP} above. This is, in fact, the case here since 
the likelihood for the particle physics model is
unaffected by the stellar kinematic data. It is also the case that the 
halo density profile and stellar velocity dispersion anisotropy parameters are
not affected by the particle physics constraints, hence the two likelihoods are independent. On the other hand, if we had included, for example, 
gamma-ray flux upper limits from Imaging Air Cerenkov Telescopes (ACTs)~\cite{Buckley:2008zc} in the particle physics likelihood, 
this would couple the two separate parameter
spaces and invalidate the above decomposition. At present including such data is not
necessary because the upper limits are well above the CMSSM parameter space, however it will be desirable and indeed necessary in the future. 

To obtain a Markov chain on the dSph model parameters, we opted for a combination of the
slice sampling \citep{Neal00} and the Metropolis-Hastings \citep{Hastings70} algorithm. The advantage of slice sampling 
is that, unlike the Metropolis-Hastings algorithm, its efficiency is not strongly linked to the proposal pdf.  
Whereas, with a good proposal pdf, the Metropolis-Hastings algorithm converges faster to the desired
distribution.  Thus, we obtain an initial proposal pdf using slice sampling and then employed it in the Metropolis-Hastings
algorithm to derive our final posterior. 
(For the actual slice sampling methodology, see \citet{Neal00} and \citet{LewisBridle06-notes}.
For the Metropolis-Hastings methodology see \citet{Christensen01, LewisBridle02, Baltz06}, and 
\citet{LewisBridle06-notes} for details).  

In practice, we found that our slice sampling run took 3-4 likelihood evaluations per point and offered
fairly good convergence.  However, our subsequent Metropolis-Hastings run had an acceptance rate
of 30\%-50\% with excellent convergence.
Nine chains of 30,000 points were
obtained, thinned, and then combined with the~\SB~ 
chains (as per the method outlined in \ref{appendix:combine}).
We refer to~\ref{appendix:convergence}
for a more detailed discussion of the convergence criteria we use for our 
chains. 

\section{Boost from Halo Substructure
\label{section:boost}} 
In the previous sections, we have outlined our modeling of the
dwarf halos from stellar kinematics and our method for scanning
the CMSSM parameter space and used this modeling to predict
the flux under the assumption of a smooth halo. 
The final ingredient we must add 
to the flux predictions is the boost from halo substructure. 
The goal of this
section is to derive the probability distribution for the boost
factor, accounting in the most reasonable possible manner
for the astrophysical and particle dark matter uncertainties
that enter the calculation. 

\subsection{Defining Boost} 
Dark matter halos form hierarchically, and this results
in a population of surviving gravitationally-bound 
substructure. High
resolution dissipationless numerical simulations
reveal substructures in $z=0$ Milky Way-size halos 
with a mass spectrum
that rises towards smaller masses with $N(>m) \propto m^{-\alpha}$ and
$\alpha \sim 0.9$, down to the smallest masses that
can be resolved, currently  
$m \gtrsim 10^{-6} M_{\rm host} \sim 10^6 M_\odot$
 ~~\citep{Diemand:2008in,Springel:2008cc}.   However, as we discuss below,
this mass is some ten orders of magnitude larger than the minimum
mass expected for CDM structure, $m_{\rm min}$.
While numerical simulations that focus on nested regions within
very high redshift halos have demonstrated
that halos with masses close to the filtering mass 
do survive the initial process of halo formation
~\citep{Diemand:2005vz,Diemand:2006ey}, more modeling
will be required to better understand the survival probability, density structure,
and precise mass spectrum of the smallest CDM substructures at $z=0$.

Because substructures themselves were assembled from smaller units prior to 
infall into their host, we expect a hierarchy of substructures within substructures 
that extends down to $m_{\rm min}$.  The first explorations
of this hierarchy of mass functions and substructure distributions is just now
becoming viable in state-of-the art simulations ~\cite{Springel:2008cc}. 
The dark matter halos of dSphs are substructures of halos of the Milky Way,
therefore their substructure fractions and mass functions are less well
explored, but there are clear expectations.   
Relative to substructures within the Milky Way halo, substructures
within dSph halos are expected to be older, and this 
gives them more time to assimilate
into the smooth component of their hosts.
Moreover, the halos of dwarf satellites do not get replenished by accreted field halos
to the extent that an isolated field halo would.   Both of these effects will act to reduce
substructure fractions in dSph halos relative to isolated galaxy halos.

Following previous treatments, we define the boost $B$ such that total gamma-ray flux $\Phi$
from a halo of mass $M$ is $\Phi(M) = 
\left[1 + B(M, m_{\rm min})\right] \tilde{\Phi}(M) $.  Here 
$\tilde{\Phi}$ is the flux that comes from a {\em smooth} halo of mass $M$
and and the boost includes a contribution from all subhalos larger than $m_{\rm min}$, which is
set by particle physics. We begin with the formulation in \citet{Strigari:2006rd}, but note that the formalism of \citet{PieriBertone08} provides similar results.   
Adopting the above definition, we may write the boost as an integral that accounts 
for substructures going down the CDM mass hierarchy~\citep{Strigari:2006rd}:
\begin{eqnarray}
{\tilde{\Phi}(M)} B(M, m_{\rm min})  = A M^\alpha
\int_{m_{\rm min}}^{qM} \left[1+B(m, m_{\rm min})\right]\tilde{\Phi}(m)
m^{-1-\alpha} dm.
\label{eq:boost}
\end{eqnarray}
Here we have assumed that halos of mass $M$ host 
substructures of mass distribution  $dN/d\ln m = A(M/m)^{\alpha}$ for $m<qM$ and (for now)
we assume a self-similar substructure hierarchy.  Written in this way, we can see that
the total boost in gamma-ray luminosity depends sensitively on a competition between
the smooth flux $\tilde{\Phi}(m)$, which tends to decrease towards smaller masses, and
$dN/d\ln m$, which rises to small masses.
For halos described by NFW density 
profiles
 \cite[e.g.][]{Navarro04-cdm.halo.3,Bullock2001} with a concentration-mass relation
that follows $c(m) \propto m^{-\mu}$ with $\mu \sim 0.1$, we expect $\tilde{\Phi}(m) \propto m^{\xi}$ with
$\xi \simeq 1- 2.2 \mu$ ~\cite{Strigari:2006rd}.  Setting $q=1$, we find that the boost $B(M,m_{\rm min})$
is approximately  proportional to $(M/m_{\rm min})^{\alpha-\xi}$.  Since the ratio 
$(M/m_{\rm min}) \sim 10^{15}$ for host halos of relevance, the precise value of
 the quantity $\alpha - \xi$ ($\sim 2.2 \mu$) at the smallest mass scales
 becomes crucial in determining whether the boost is significant or negligible.
By making the most optimistic assumptions possible
($q=1$, self-similar substructure hierarchies, and optimistic assumptions for the density structures of small
halos with $\mu \sim 0.1$) ~\citet{Strigari:2006rd} and~\citet{Kuhlen:2008aw}
have shown that 
the boost should be no larger than about 100, and more typically $\lesssim 10$
 for $m_{\rm min} \sim 10^{-6} M_\odot$.

\subsection{Analytic Solution for Boost} 

The above solutions to~\Eref{eq:boost}
~\citep{Strigari:2006rd,Kuhlen:2008aw}
were obtained by numerical integration. 
Here, we present an analytic solution to~\Eref{eq:boost}, 
and discuss its utility.  

We begin by assuming that the flux at a fixed host halo mass scales 
roughly as a power law, $\tilde{\Phi}(M) \propto M^{\xi}$.  With
this, we may rewrite~\Eref{eq:boost} as
\begin{equation}
\label{eq:dboost}
D(b)' = A\left[q^{\xi - \alpha}\exp(b) + D(\ln q + b)\right]
\end{equation}
where $b = \ln q + \ln M - \ln m_{\rm min}$ and $D(M) = M^{\xi -
\alpha} B(M)$.  Note that $B(m_{\rm min}/q) = D(0) = 0$.  It should be
noted that $B(m_{\rm min}/q) = 0$ and not $B(m_{\rm min}) = 0$ is the appropriate
boundary condition (as was assumed in \citet{Strigari:2006rd}).  
Using these boundary conditions 
~\Eref{eq:dboost} can be recursively solved as:
\begin{equation}
\label{eq:bnrecur}
B(M, m_{\rm min}) = 
\cases{
0 & $M \le m_{\rm min}/q$ \\
A q^{\xi - \alpha}
\frac{\left(\frac{q M}{m_{\rm min}}\right)^{\alpha-\xi}-1}{\alpha-\xi} 
& $m_{\rm min}/q < M \le m_{\rm min}/q^2$\\
\quad \vdots & $\quad \vdots$\\
}
\end{equation}
This forms a set of functions, $B(M, m_{\rm min}) = \{B_0(M, m_{\rm min}),
B_1(M, m_{\rm min}), \dots, B_n(M, m_{\rm min})\}$ where each function $B_n(M,
m_{\rm min})$ is only valid in the interval $m_{\rm min}/q^{n} < M \le 
m_{\rm min}/q^{n+1}$. 

Conceptually, the $B_n$'s represent the amount of substructure included in the
calculation of the boost.  For example, $B_0(M, m_{\rm min})$ represents
the boost of a halo with no substructure, and thus we set 
$B_0(M, m_{\rm min}) = 0$.  $B_1(M, m_{\rm min})$
represents the boost with the inclusion of only the subhalos whereas
$B_2(M, m_{\rm min})$ includes only the subhalos and sub-subhalos.
$B_n(M, m_{\rm min})$ now can be related through~\Eref{eq:bnrecur} by
\begin{equation}
\label{eq:dnboost}
D_{n}(b)' = A\left[q^{\xi - \alpha}\exp(b) + D_{n-1}(\ln q + b)\right].
\end{equation}
We may now solve for $B_n(M, m_{\rm min})$ by taking the Laplace transform
of~\Eref{eq:dnboost} (see \ref{appendix:boost}).  After inversion,
this yields (for $n > 0$)
\begin{equation}
\label{eq:boostans}
B_n(M, m_{\rm min}) = \sum_{i=1}^{n} \frac{1}{(i-1)!} 
\left(\frac{A q^{\xi - \alpha}}{\xi - \alpha}\right)^{i}\gamma 
\left(i, (\xi-\alpha)\ln\left(\frac{q^{i} M}{m_{\rm min}}\right)\right)
\end{equation}
where $\gamma(a, x)$ is the lower incomplete gamma function defined by
\begin{equation}
\gamma(a, x) = \int_0^x x^{a-1} \exp(-x) d x.
\end{equation}

In the above analysis, we have assumed that the mass function
is self-similar for all levels of substructure. However, this is unlikely
to be the case and simulations~~\citep{Diemand:2008in,Springel:2008cc} see less substructure 
in subhalos than in host halos. To account for the fact that the mass 
function may differ for each level of substructure, 
it would be useful to perform the same analysis for the case where the
mass function varies independently for each level of substructure.
We define the mass function to be
\begin{equation}
\label{eq:subhalomassfuni}
\frac{d N}{d \ln m} = A_i \left(\frac{M}{m}\right)^{\alpha_i}
\end{equation}
at level $i$.  Here, $i=0$ would apply to the parent halo whereas
$i=1$ would apply to the subhalos and so on.  Also, we let 
$q \rightarrow q_i$ following the same notation.  
Using the same analysis
as above, we can rewrite~\Eref{eq:boostans} as
\begin{equation}
B_n(M, m_{\rm min}) = \sum_{i=1}^{n} \frac{1}{(i-1)!} \left(\frac{\tilde{A}_i 
\tilde{q}_i^{\xi - \alpha_i}}{(\xi - \alpha_i)^i}\right)\gamma 
\left(i, (\xi-\alpha_i)\ln\left(\frac{\tilde{q}_{i} M}{m_{\rm min}}\right)\right)
\end{equation}
where $\tilde{A}_i \equiv \prod_{j=1}^i A_j$ and $\tilde{q}_i \equiv \prod_{j=1}^i q_j$.
For completeness, we present the $\alpha_i=\xi$ solution :
\begin{equation}
B_n(M, m_{\rm min}) = \sum_{i=1}^{n} \frac{\tilde{A}_i\left(\ln\left(\frac{\tilde{q}_{i} M}{m_{\rm min}}\right)\right)^i}{i!}
\end{equation}

The utility of the analytic solution derived above stems from the fact that it lets us account for each level of substructure separately.  We find that in the most extreme circumstance of $\xi - \alpha = -0.2$ and $m_{\rm min} = 10^{-10} M_{\odot}$, inclusion of only the subhalos and sub-subhalos leads to an accuracy of $98.4\%$.  Thus, it is not necessary to go below $n=2$, for accurate boost predictions -- for most part it is sufficient to just resolve subhalos to estimate the boost. 

\subsection{Annihilation luminosity - mass relation}
\label{sec:lumfun}

In order to compute the boost, we need to estimate the gamma-ray
luminosity from a subhalo of the halo under consideration. The
luminosity depends strongly on the log-slope of the inner density
profile of the subhalo and the effect of the tidal forces of the
parent halo. In order to estimate these effects, we consider a
simplified model of subhalo distribution and tidal effects. 

We consider two possibilities for the spatial distribution
of the subhalos. One is motivated by the current 
high-resolution simulations of Milky Way sized halos 
~\citep{Diemand04,Gao07,Madau08,Springel08}, where it
was found that the number density of subhalos with masses greater
than $\sim 10^{-6}$ the mass of the parent halo falls off as $1/r^2$ at large
radii. At small radii, tidal forces of the parent halo will destroy
the subhalo and hence reduce the population in the inner most
regions. However, it is not clear that the smallest subhalos extending
down to earth mass subhalos should have this, more diffuse, spatial
distribution. We therefore also consider the other extreme where the
spatial distribution of the subhalos follows the smooth
distribution. We choose the second possibility for the final
computations because the boost is primary affected by the smallest
mass halos. We randomly generate subhalos according to the chosen
spatial distributions. 

Before it fell into the host halo, 
we allow for the log-slope of the subhalo inner density profile to range 
between $a=0.8$ and $a=1.4$, motivated
~\citep{Diemand04-cusps}. We note, however, that more recent simulations indicate that there is no
asymptotic inner slope~\citep{Navarro04-cdm.halo.3, Springel:2008cc,Navarro:2008kc}. The density profile
assumed for the subhalo when it was outside the host halo is
\begin{equation}\label{eq:boost-calc-density}
\rho_a (r) = \rho_0 (r/r_0)^{-a} (1+r/r_0)^{-3+a}\,,
\end{equation}
where $a=1$ corresponds to the NFW profile and the concentration is defined by the ratio of the
virial radius and the scale radius as $c = R_{\rm vir}/r_{0}$.
We assume a power-law virial mass function for the subhalo before it
fell into the host halo and generate masses randomly for the
subhalos. Since we start by picking the virial mass of the halos, we
use the observed mass function for field halos. However, 
there is a huge extrapolation involved -- we assume that the power-law
extends all the way down to the minimum halo mass where the mass
function gets truncated.  Given the virial mass, we find the
concentration of the halo using the field concentration--mass
relation including the scatter. While these relations were derived for
the NFW profile, we adapt them for the profile in
~\Eref{eq:boost-calc-density}. Assigning a concentration is the most
uncertain part of the calculation; in fact, as we will see, {\em the flux
  depends sensitively on the concentration of the smallest halos.}

Following the above procedure, we are able to set $r_0$ and $\rho_0$, 
and from this we then find the
corresponding ergodic distribution function, $f_a(E)$. Given the 
implied tidal radius, we adopt the following simple model for tidal effects
wherein any given tidal radius defines a relative energy $E_0$ below
which the distribution function of the subhalo drops to zero. This is
analogous to the King models for the stars. That is, 
\begin{eqnarray} 
f(E) & = & f_a(E) \, \forall \, E>E_0 \,, \label{eq:tidalfe} \\
     & = & 0 \, \forall \, E < E_0 \nonumber \,. \\
\end{eqnarray}
Two points are worth noting: (1) $E_0$ is fully specified by the
ratio of the tidal radius to $r_0$, and (2) the product
$\rho_0 r_0^a$ has to be unchanged because the tidal effects do
not change the innermost regions of the subhalo. We find that
the density profiles resulting from the distribution function in
~\Eref{eq:tidalfe} are well fit by 
\begin{equation}
\rho_{\rm sub} (r) = \rho_a (r) \exp(-(r/r_t)^n)\,,
\end{equation}
where $r_t$ is defined here as the tidal radius and $n$ is a function
of $E_0$ (or equivalently $r_t/r_0$). For any given $\rho_0$ and
$r_0$, this model then provides a relation between $r_t$ 
and the tidally truncated mass of the
subhalo. If $r_t$ is smaller than $r_0$, we consider the subhalo
destroyed. 

To fix the mass and tidal radius for a given $\rho_0$, $r_0$ and
distance from the center of the host halo, we iteratively solve the 
Jacobi relation between $r_t$ and mass, under the approximation that 
both the  host halo and the subhalo are isothermal. We allow for the
fact that the satellite could have been closer in its orbit by
choosing the closest distance to the center along the orbit to be a
factor less than one, chosen randomly from 0 to 1, times the current
distance.   

This is now a fully specified model and we may now predict the
gamma-ray annihilation luminosity as a function of the tidally
truncated mass of the subhalo. At the high mass end of the subhalo mass
function, the predictions will have large scatter because of sample 
variance. At smaller masses, where most of the flux comes
from, the large number of subhalos contributing to the flux means that
the predictions essentially have no scatter. 

As a means of illustrating the important uncertainties associated with
subhalo density structures we explore two simple 
models for the concentration-mass relation.  The first is a 
power law (PL) model $c(m) \propto m^{-0.1}$ that is normalized to match the
results of N-body simulations over the mass ranges that have been
probed by direct simulations ($m \gtrsim 10^{8} M_\odot$).
The second is the simple analytic model~\citet{Bullock2001} with the
specific implementation of~\citet{Maccio:2008xb}.
This model (denoted B01 below) also matches simulations down to the mass-scales
explored by simulations, but links the value of $c$ to the
spectrum of CDM density fluctuations via estimated collapse times~\footnote{The choice of B01 is representative of many
proposed analytic models that link $c$ to the power spectrum.  Among these it has
the {\em steepest} faint-end slope.}.  While the
power-law extrapolation leads to concentrations of order 500 
for earth-mass halos, the B01 model concentrations are much lower,
and plateau for small masses because the slope of the power spectrum of density
fluctuations is more shallow on small scales. 

We note that the flux depends primarily on the profile around $r_0$
and smaller. The concentration sets $\rho_0$ since $\rho_0 \propto
c^3$ and the flux is proportional to $\rho_0^2 r_0^3$.  
We find that the scaling formula suggested by~\citet{Strigari:2006rd},
$\tilde{\Phi}(M) \propto M\,c(M)^{2.2}$, works well (here $M$ is the subhalo
tidally-truncated mass) for power-law $c(M)$ functions. We did not find
any systematic differences in this scaling for the two assumptions
about the spatial distribution of the subhalos. However, we do find
that the more concentrated spatial distribution (NFW) implies
systematically larger fluxes (by a factor of about 2). In the
self-similar calculation below we only include the scaling
information. 

We do not consider the effect of the assumptions about
the spatial distribution of subhalos on the signal-to-noise. In
particular, if the distribution of the subhalos is as shallow as
$1/(r_s+r)^2$, then much of the signal (even for more moderate boosts of
order 1) will come from the ``outer'' regions. As we integrate the
signal outwards from the stellar core, two effects must be considered.
One, the signal-to-noise depends sensitively on the angular acceptance
region around the satellite. If the background is mainly extragalactic
(or at least constant across the angular region of the galaxy), 
it will scale linearly with
the angular region covered. The scaling of the signal with the angular
acceptance can only be estimated if the spatial distribution of the 
subhalos of a satellite is known. Two, we cannot accurately predict
the tidal radius beyond which both the dark matter density profile of
the satellite galaxy is cut-off. For the
boost calculation in the MCMC exploration, we assume that the spatial
distribution of the subhalos follows that of the smooth halo. We
reiterate that the relevant quantity is the distribution of small
subhalos that cannot yet be resolved by simulations and those
subhalos are the ones that contribute dominantly to the gamma-ray
flux. 

\subsection{Minimum Mass CDM Halo} 
\label{sec:mmin}

We now have an expression for the boost as a function of (1) the host
halo mass, (2) the mass function of CDM subhalos, (3) the concentration-mass
relationship for CDM subhalos,  (4) the 
minimum mass CDM halo, $m_{\rm min}$. The uncertainty in the fourth item
arises from the unknown cut-off scale in 
the mass function of CDM substructure. As mentioned above,
this cut-off scale is well below the resolution of modern day 
CDM numerical simulations of Milky Way like galaxies at $z=0$.  
The smallest halo size is set by the horizon at kinetic decoupling
in the early universe. The kinetic decoupling in turn depends on the
scattering interactions of dark matter with standard model 
fermions, as well as the free-streaming after decoupling
~\citep{Loeb:2005pm,Profumo:2006bv,Bertschinger06,Bringmann:2006mu}, 
\begin{equation}
m_{\rm min} = 5.72\times10^{-2} \Omega_M h^2 C^{3/4}
\left(\frac{m_{\chi}\sqrt{g_{eff}}}{100 GeV}\right)^{-15/4} M_{\odot}. 
\label{eq:mcdm}
\end{equation}
Here $g_{eff} = 10.75$ is the effective degrees of freedom when 
the CDM particle freezes out, and 
\begin{equation}
C \equiv \frac{m_{\chi}^2}{p_{cm}^2}\left|\mathcal{M}(t \rightarrow 0)\right|^2,
\label{eq:scatteringmatrix}
\end{equation}
where $\left|\mathcal{M}(t \rightarrow 0)\right|$ is the matrix
element of elastic scattering 
in the limit of no momentum transfer ($t \rightarrow 0$).  

\citet{Bertschinger06} has calculated $m_{\rm min}$ assuming 
a Bino-like neutralino, finding typical values $m_{\rm min} \sim 10^{-4}$
M$_\odot$ for typical WIMP masses. We follow the kinetic decoupling calculation of~\citet{Bertschinger06} but generalize it to include Wino and Higgsino-type neutralinos by determining $C$ at each point in the~\SB~ chain. 
In~\ref{appendix:scat},  we present the relevant
Feynman rules and matrix elements that enter into calculating
the scattering matrix in~\Eref{eq:scatteringmatrix}. 

It is important to note that in our analysis of the scattering matrix 
element, we assume that before kinetic decoupling occurs, 
the WIMP interacts solely with electrons and neutrinos. In doing
so, we have neglected the effects of muon scattering, which will
be relevant if kinetic decoupling occurs at temperatures comparable 
to the muon mass. Including the non-negligible muon abundance
would in turn modify~\Eref{eq:mcdm}, which is beyond the scope
of our present analysis. 

\begin{figure}
\begin{center}
\rotatebox{270}{\includegraphics[height=0.482\hsize]{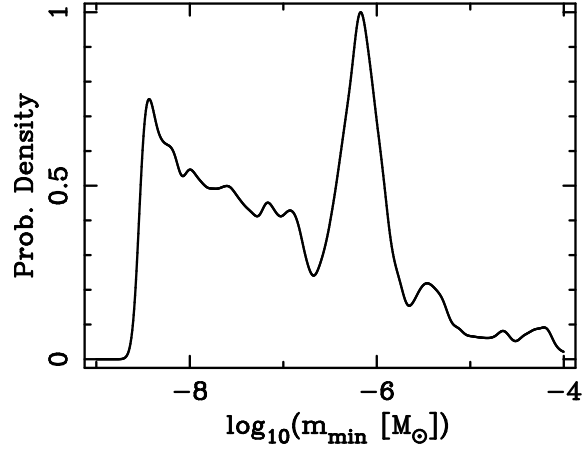}}
\caption{\label{fig:m0}\footnotesize 
Posterior probability for m$_{\rm min}$ in the CMSSM, assuming flat 
priors and employing all available constraints.
}
\end{center}
\end{figure}

In \Fref{fig:m0}, we show the resulting posterior distribution of $m_{\rm min}$ 
accounting for the entire presently viable parameter space of the CMSSM.  
We find $\sim 95\%$ c.l. values for the minimum halo mass 
within the range $\sim10^{-9} - 10^{-6}$ M$_\odot$.  
The features in the likelihood for $m_{\rm min}$ arise from the
probability distributions in the CMSSM parameters; due to the
non-linear transformation between these parameters and 
the minimum mass halo in~\Eref{eq:mcdm} any small features
in the CMSSM parameters are strongly exacerbated in 
the $m_{\rm min}$ likelihood.  
\begin{figure}
\begin{center}
\rotatebox{270}{\includegraphics[height=0.482\hsize]{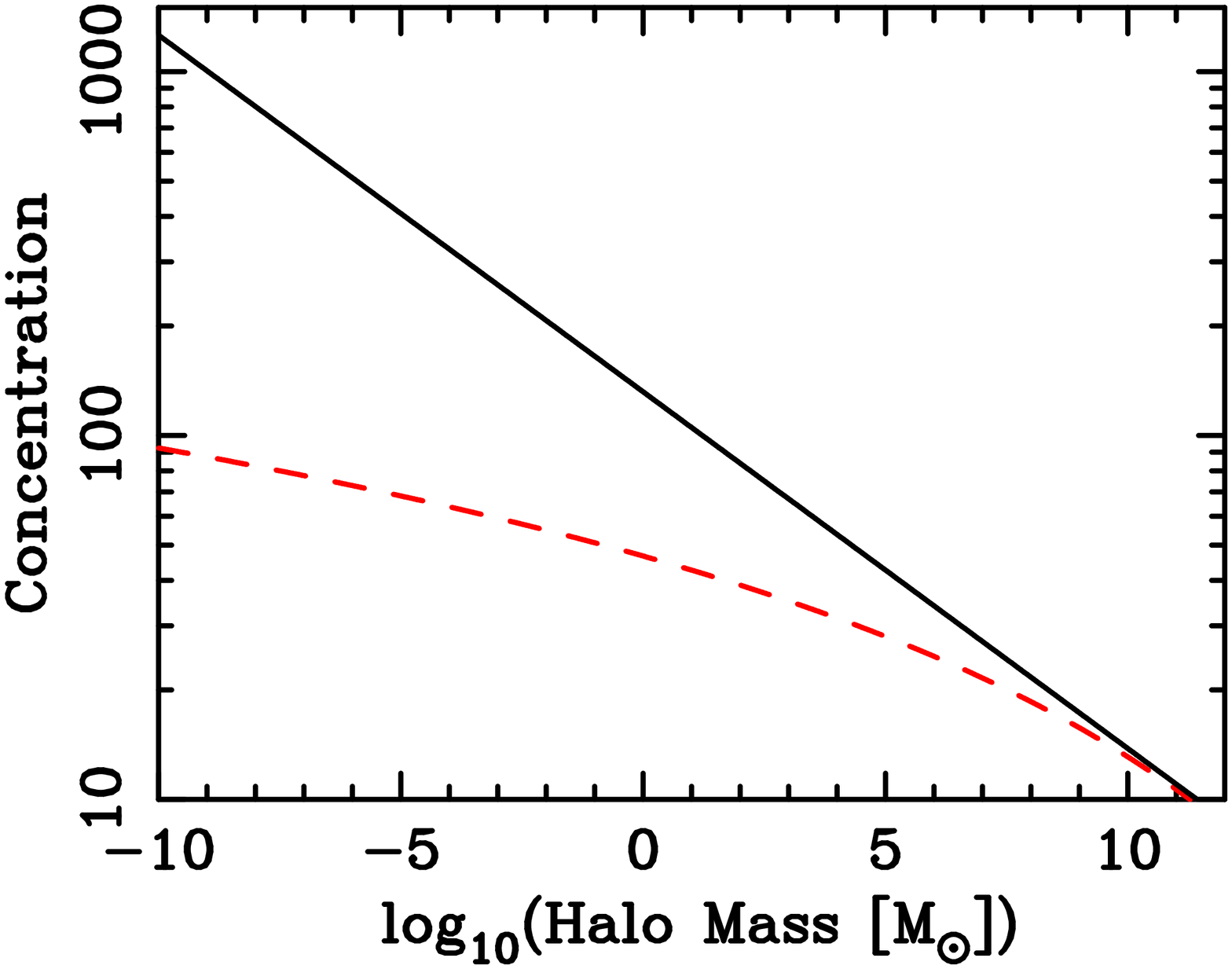}}
\rotatebox{270}{\includegraphics[height=0.482\hsize]{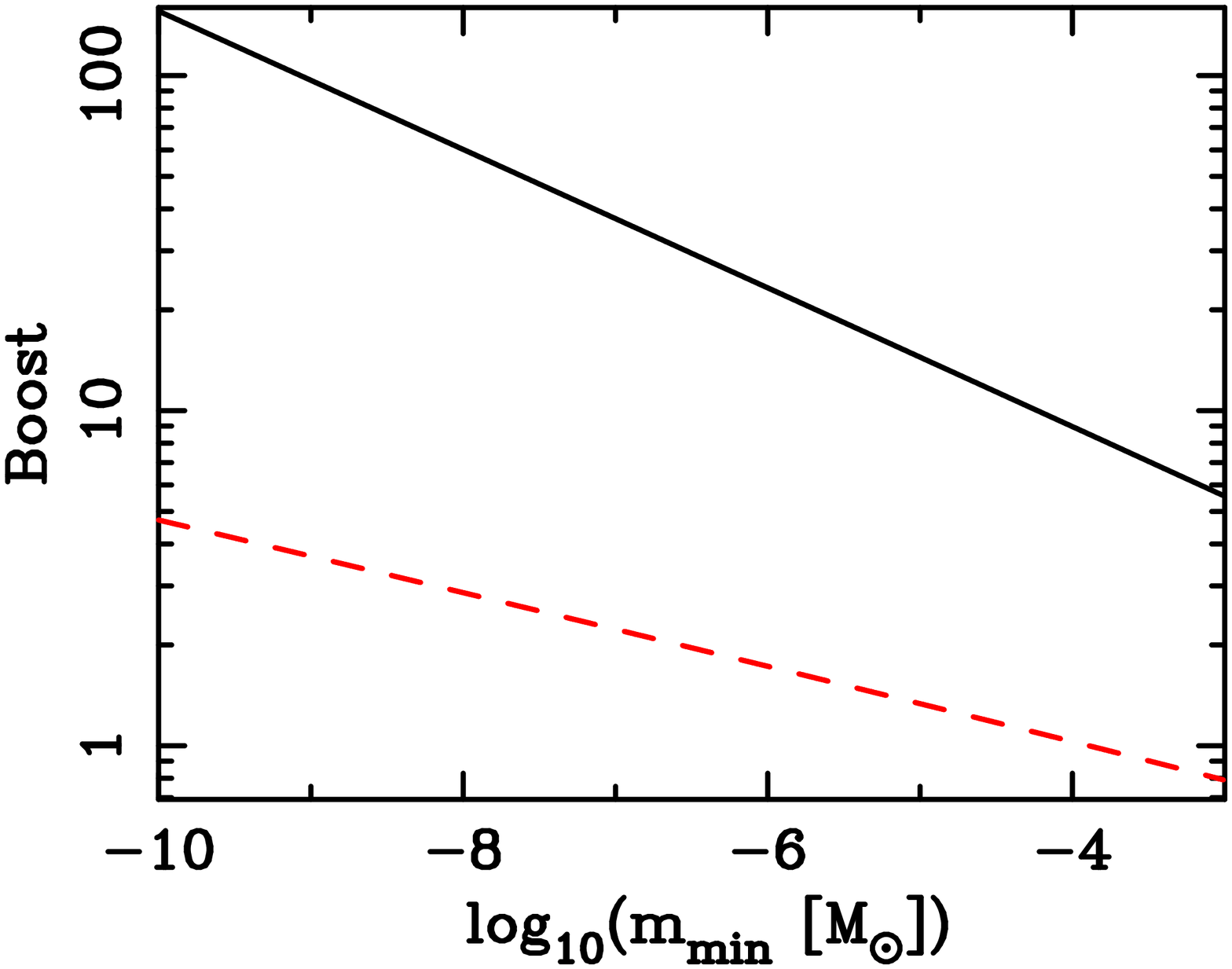}}
\caption{\label{fig:cm0}\footnotesize {\it Left}: Halo concentration
versus halo mass for our PL model (solid black line) and 
the B01 model (dashed red line).  
{\it Right}: The boost for an $M =  10^8 M_\odot$
halo that results from the PL model (solid black)
 (using~\Eref{eq:boostans}) and the B01 (dashed red) concentration models.   
Note that both of these concentration-mass models are consistent with current simulations.
}
\end{center}
\end{figure}

\subsection{Boost Predictions:  Two models \label{subsec:boostmodels}}

\begin{figure}
\begin{center}
\rotatebox{270}{\includegraphics[height=0.482\hsize]{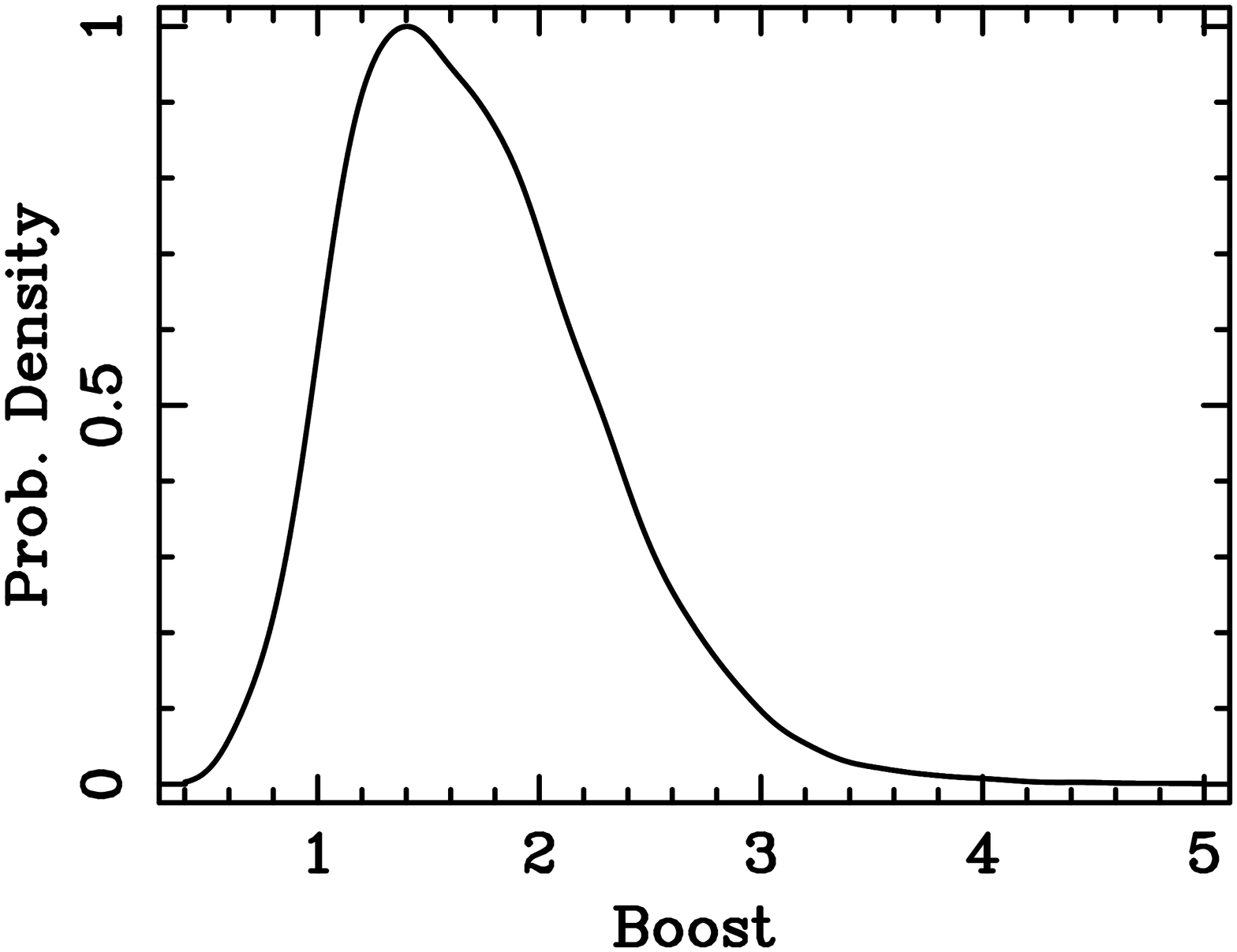}} 
\rotatebox{270}{\includegraphics[height=0.482\hsize]{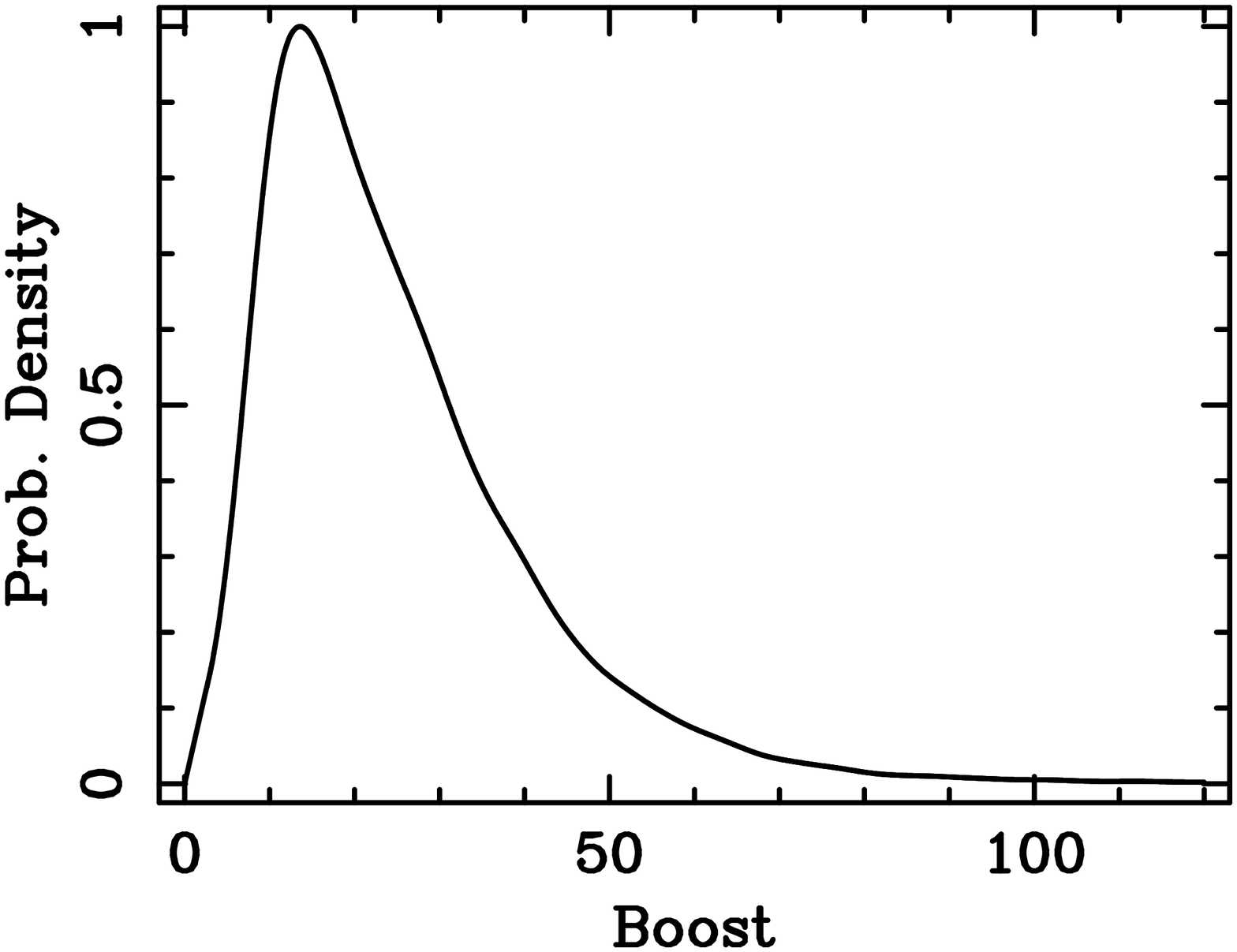}}
\caption{\label{fig:boost}\footnotesize {\it Left}: Posterior distribution
value for the boost factor, assuming the concentration-mass model of B01. 
{\it Right}: Posterior distribution  
value for the boost factor, assuming the PL concentration-mass model. 
}
\end{center}
\end{figure}

Running Monte Carlo simulations (see~\sref{sec:lumfun}),
we found that the luminosity $\Phi(M)$ has a power law behavior for both the PL and B01 models.
In terms of the first-order estimates of the boost discussed in association with
~\Eref{eq:boost} above, the relevant combination 
of  the mass-function log-slope ($-\alpha$) and the
luminosity function log-slope ($\xi$) takes the value
 $\alpha -\xi \simeq 0.2$ and $0.1$, for the PL and B01 models, respectively.  
\Fref{fig:cm0} shows the boost for a $10^8$ M$_{\odot}$
halo across its relevant range for both models. The ensuing posterior pdf for the boost for 
Segue 1 for these two models is shown in \Fref{fig:boost}.  The difference in boost 
between these two models is an  order of magnitude and this
underscores the large effect that the (uncertain) concentration of the lowest mass halos has on 
the overall boost.  This leads to a natural uncertainty for any flux prediction and emphasizes the need
to measure the concentration and mass function well for halos of the smallest size.
In order to be conservative,  for the remainder of this paper we assume $\alpha - \xi = 0.1$.

\section{Flux Predictions and Detection Prospects} 
\label{section:results}
\begin{figure}
\begin{center}
\rotatebox{360}{\includegraphics[height=0.45\hsize]{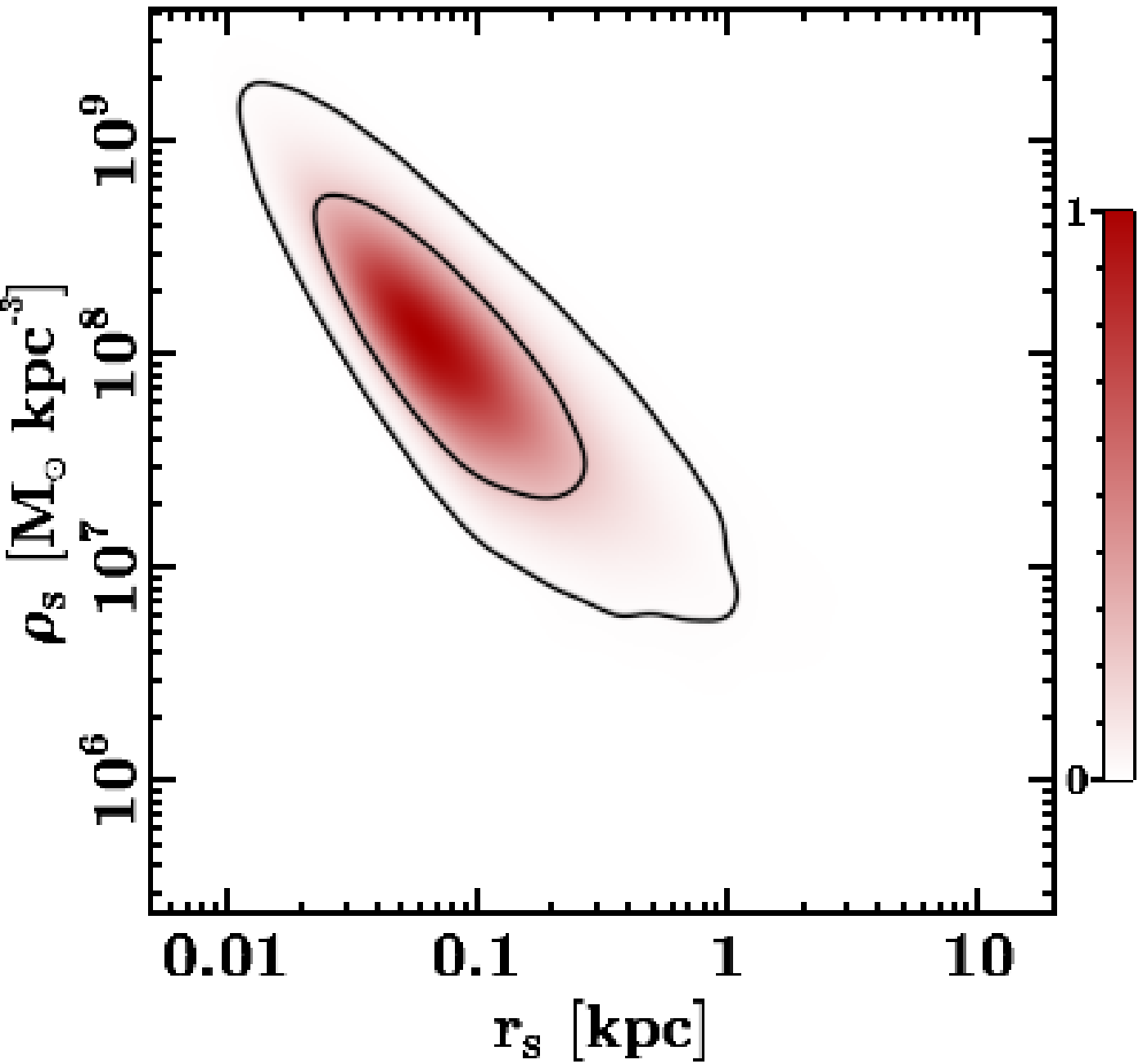}}
\rotatebox{360}{\includegraphics[height=0.45\hsize]{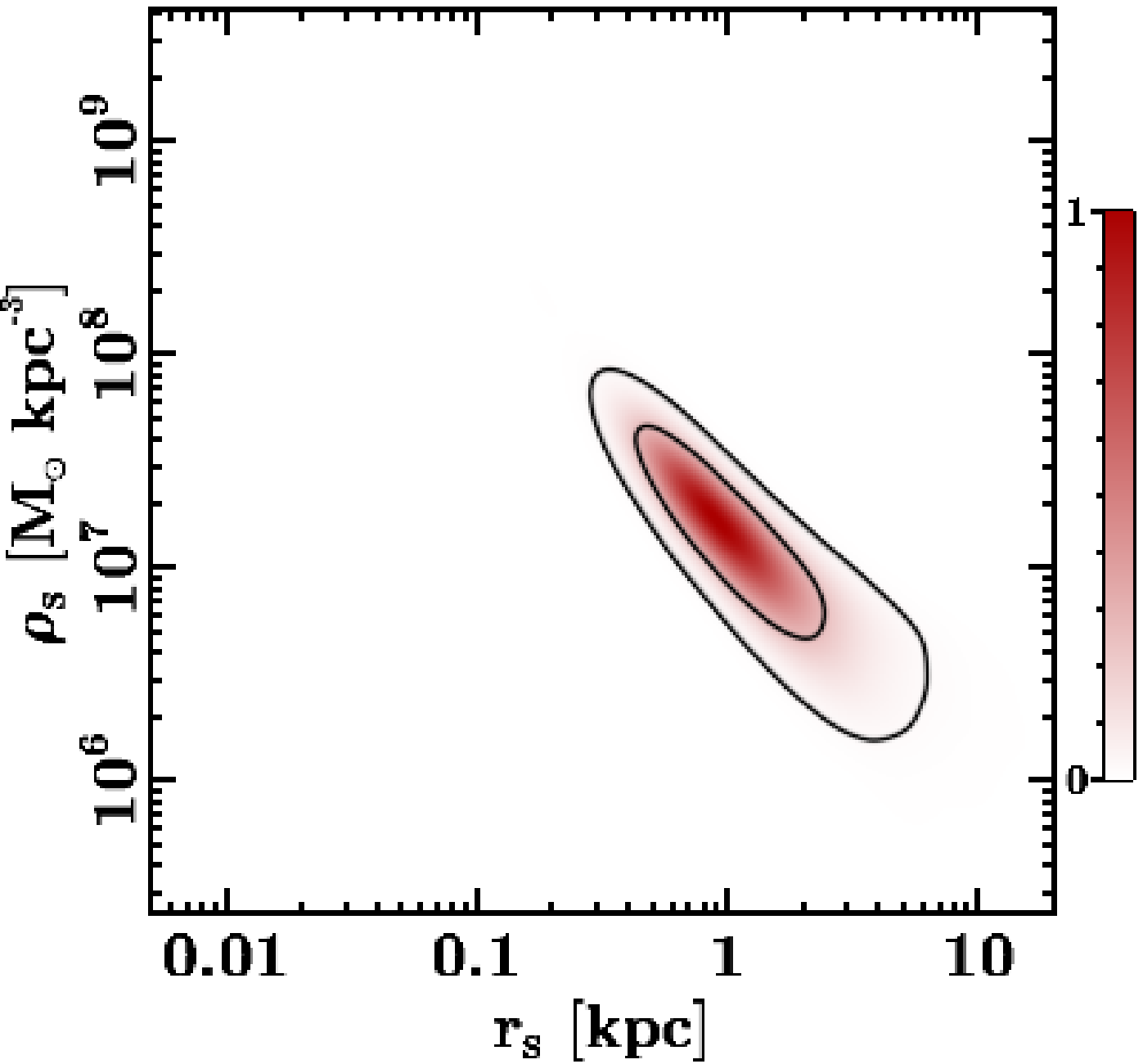}}
\caption{\label{fig:rhosrs}\footnotesize
Allowed parameter space for the  scale radius and scale density for the 
Einasto profile in~\Eref{eq:einasto} for Segue I ({\em Left}) and 
Draco ({\em Right}). 
Inner and outer contours represent the 68\% and 95\% c.l. regions
respectively.  
}
\end{center} 
\end{figure}

In the analysis above we have assembled the necessary ingredients to make
gamma-ray flux predictions given constraints from the halo
kinematics and the CMSSM. 
The corresponding kinematic constraints on the (smooth) dark matter halos of Draco and Segue 1 are illustrated in
~\Fref{fig:rhosrs}. Now, as a final step in the process
of making the flux predictions, we must specify the angular region
around the center of the dSph within which the flux is calculated, 
.i.e. $\theta_{\rm max}$ in~\Eref{eq:flux}. Determining the most optimal value 
of $\theta_{\rm max}$ 
which maximizes the signal and minimizes the background contribution
depends on the experiment and backgrounds under consideration. 
Satellite experiments such as Fermi~\cite{Atwood:2009ez} and ACTs, such as HESS~\cite{Aharonian:2007km} and
MAGIC~\citep{SanchezConde:2009ms}, are expected to have an angular 
resolution of $\sim 0.2^\circ$ and $0.1^\circ$, respectively.
If the extent of the gamma-ray emission region for the dSphs is larger than
these point-spread functions the dSphs would be resolved as point
sources.  

Our approach in this paper is to provide an algorithm for determining the
optimal value of $\theta_{\rm max}$, given the best fitting halo parameters
and the measured background spectra. More specifically, we determine the 
optimal value of $\theta_{\rm max}$ from our Markov chains, along with the
estimation of the detection significance, $\sigma = N_s/\sqrt{N_s+N_b}$, where
$N_b$ is the number of background photons and $N_s$ is the number of 
signal photons within a given angular acceptance. We note that the definition
of detection significance here is only meant to be an approximation; more 
realistically the maximum signal-to-noise will depend on the angular 
distribution of the signal and background, and detailed detector specifications. 
Nonetheless, proceeding with the above definition, at each point in our
chains, we calculate the value of $\theta_{\rm max}$ for which the significance
$\sigma$ is maximized, and then construct the pdf of the quantity, 
$\theta_{\rm max}$. In order to convert the flux to a detected number of photons, 
we take the total exposure, defined as the orbit-averaged effective area times
the observation time, to be 
$3 \times 10^{11} {\rm cm}^2 {\rm s} \simeq 2000 {\rm cm}^2 \times 5 {\rm years} $~\cite{Atwood:2009ez}. 

\begin{figure}
\begin{center}
\rotatebox{270}{\includegraphics[height=0.45\hsize]{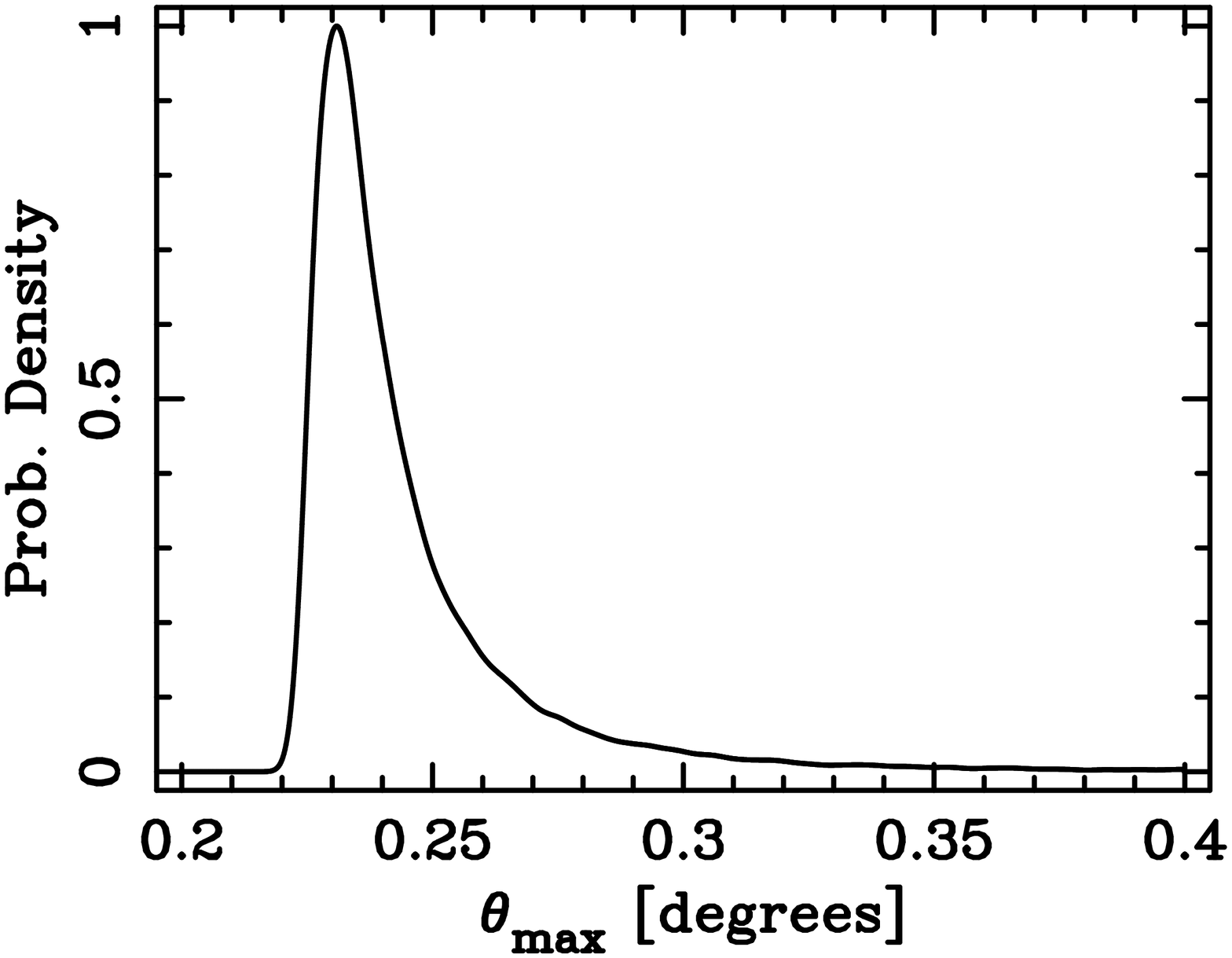}}
\rotatebox{270}{\includegraphics[height=0.45\hsize]{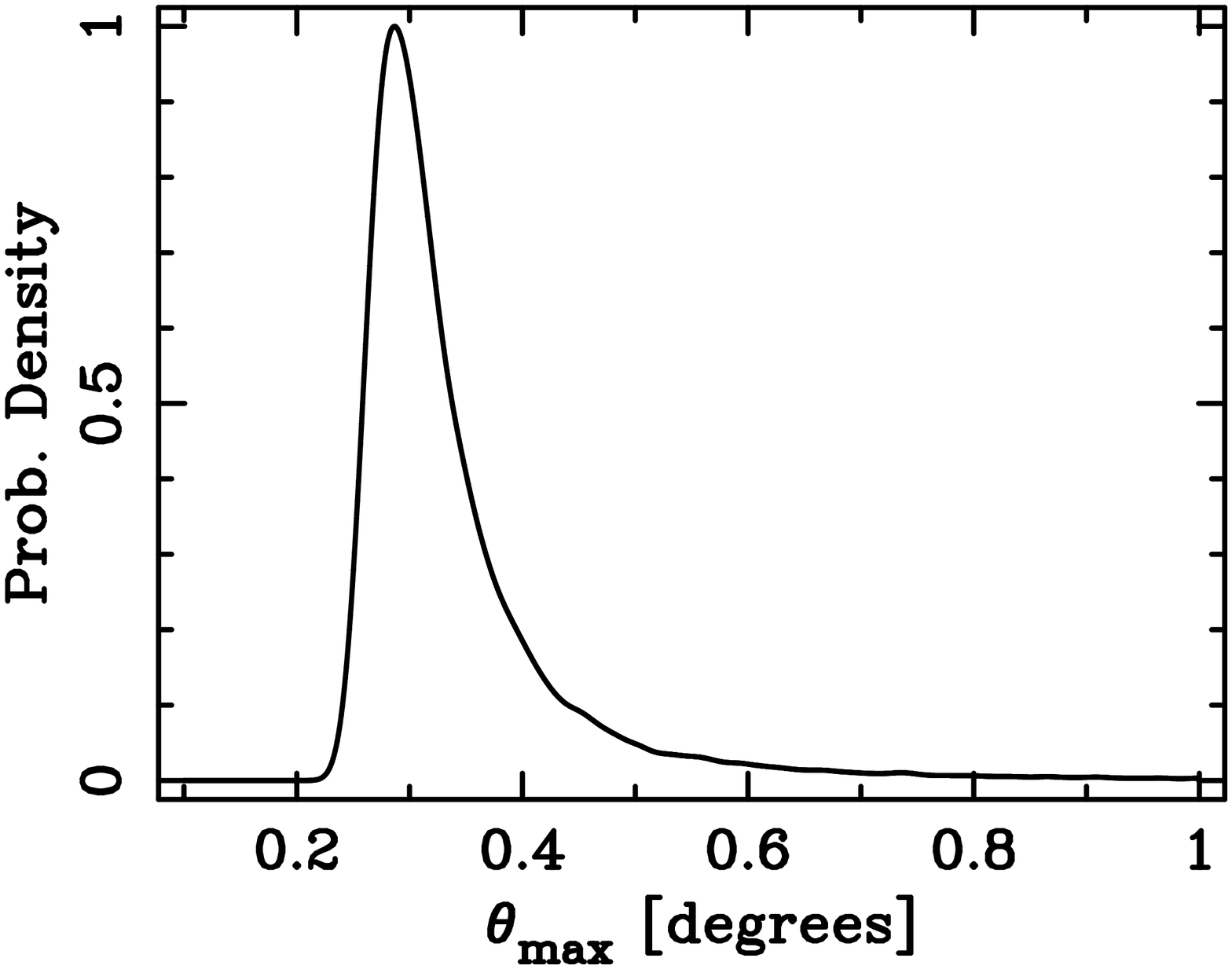}}
\caption{\label{fig:thetamax}\footnotesize
The posterior probability distribution for the angular region within which
the signal-to-noise ratio is maximized. {\em Left} panel is for Segue 1, 
and {\em Right} panel is for Draco. 
The sharp cut-off at low $\theta_{\rm max}$ 
is a result of the assumed point-spread function. 
}
\end{center} 
\end{figure}

For our input background spectrum  
we perform a standard analysis utilizing the EGRET diffuse background
measurements at high Galactic latitude. We consider just the diffuse extragalactic
background, though note that including contributions from the residual Galactic
component may increase the total background by flux about an order of magnitude. 
The diffuse backgrounds over
the relevant energy range of $\sim 1-100$ GeV for WIMP annihilation will of course be mapped
with even greater precision in the very near future with Fermi.
The diffuse extragalactic background is seen to fall off 
according to the power law $dN/dE \sim E^{-2.1}$~\cite{Hunter1997}; 
we integrate this background spectrum for energies $> 1$ GeV, and 
compare to the number of photons produced by dark matter over 
the same energy range. 
The resulting pdf for $\theta_{\rm max}$ is 
shown in~\Fref{fig:thetamax}. 
As~\Fref{fig:thetamax} shows, accounting for the diffuse backgrounds, the optimal 
angular extent for each galaxy is $\sim 0.2^\circ$ (the PSF of Fermi), 
with a gradual tail that extends to larger angles. 

\begin{figure}
\begin{center}
\rotatebox{0}{\includegraphics[height=0.45\hsize]{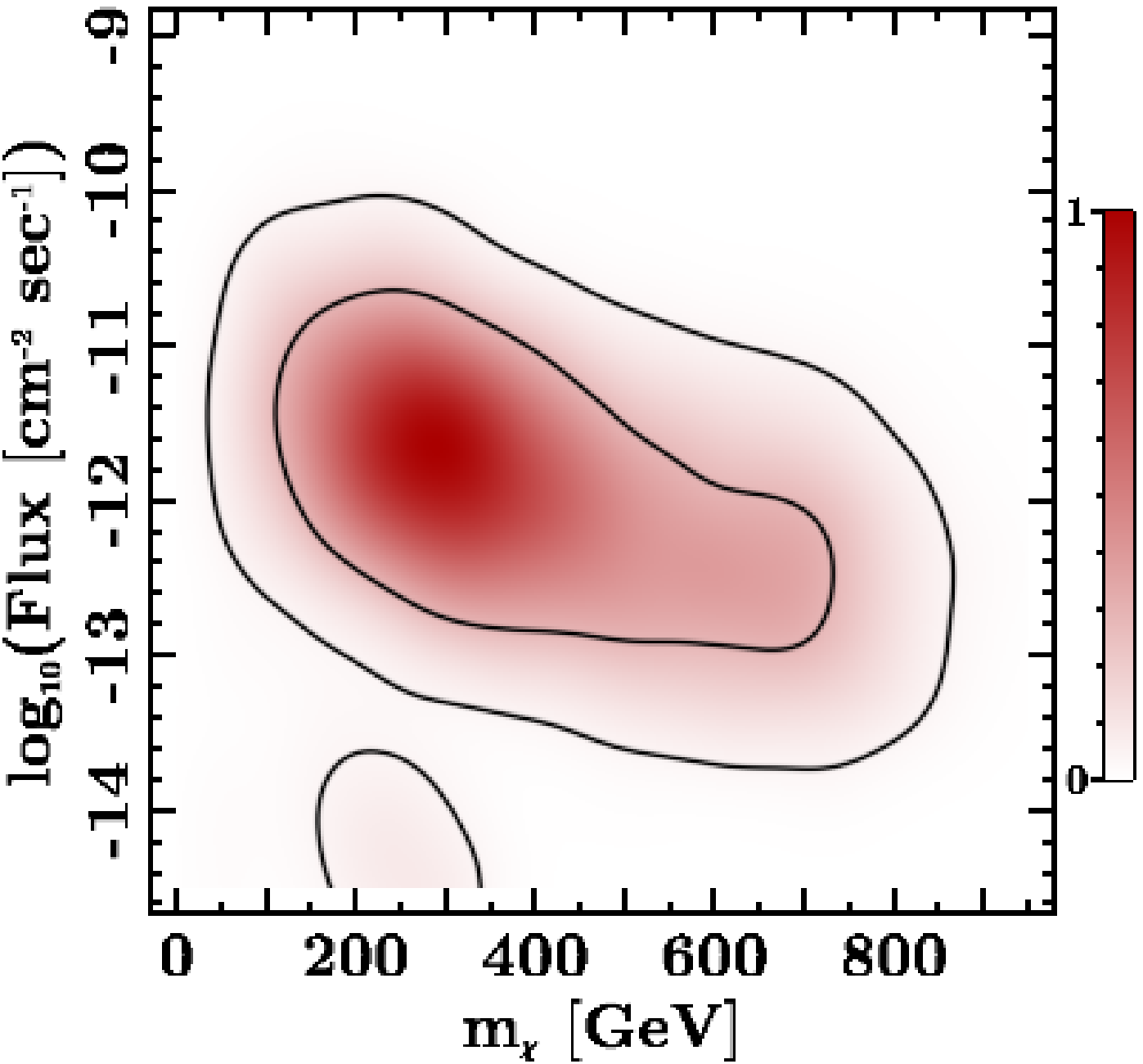}}
\rotatebox{0}{\includegraphics[height=0.45\hsize]{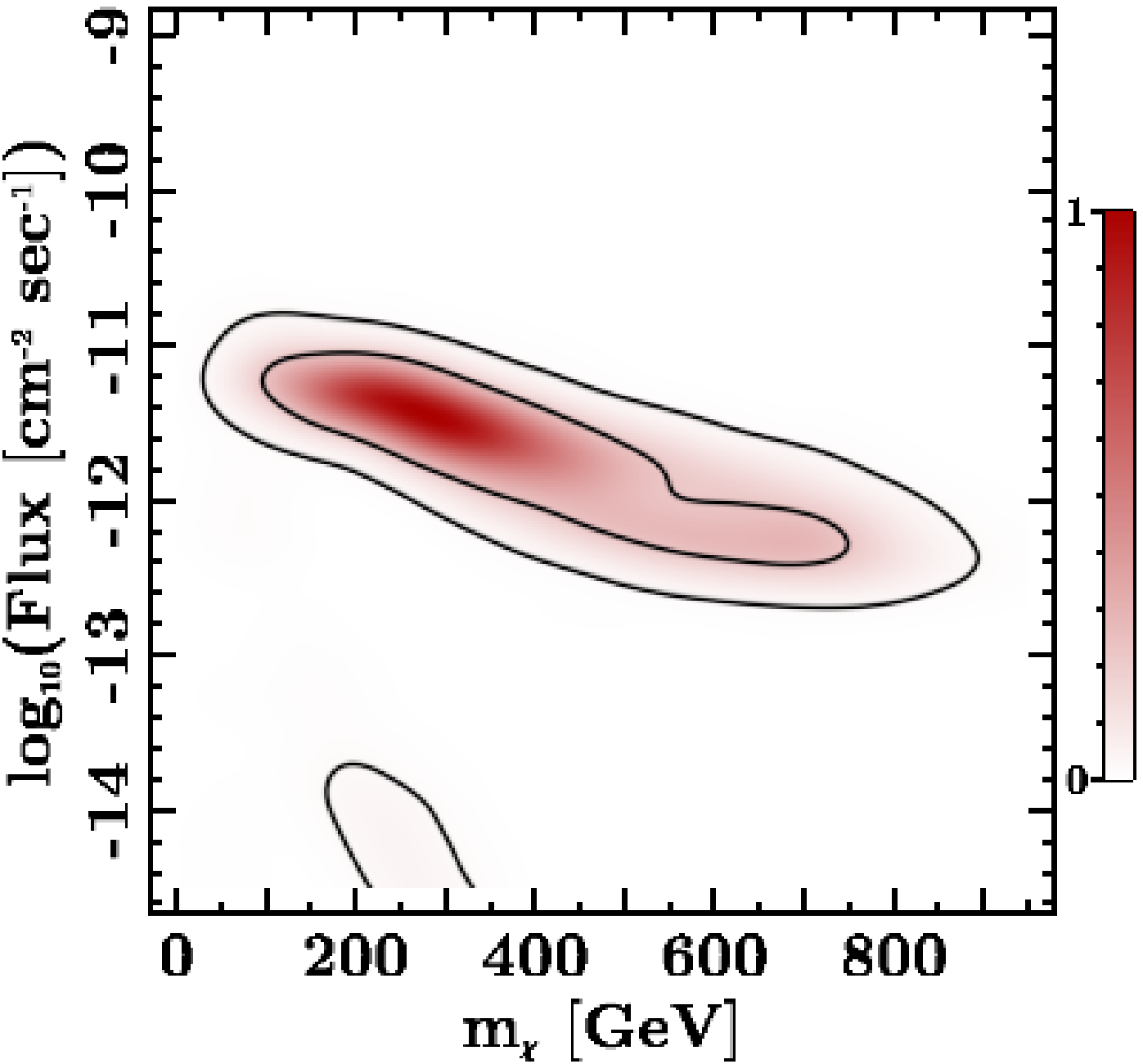}}
\caption{\label{fig:flux_mx}\footnotesize
The $E > 1$ GeV gamma-ray flux versus the neutralino mass for Segue 1 ({\em Left}) and 
Draco ({\em Right}). 
Both figures include the conservative boost model corresponding to
the~\citet{Bullock2001} halo concentration model
(discussed in~\Sref{subsec:boostmodels}) and CDM prior (discussed
in~\Sref{subsec:priors}). 
Inner and outer contours represent the 68\% and 95\% c.l. regions
respectively.  
}
\end{center} 
\end{figure}

\begin{figure}
\begin{center}
\rotatebox{0}{\includegraphics[height=0.45\hsize]{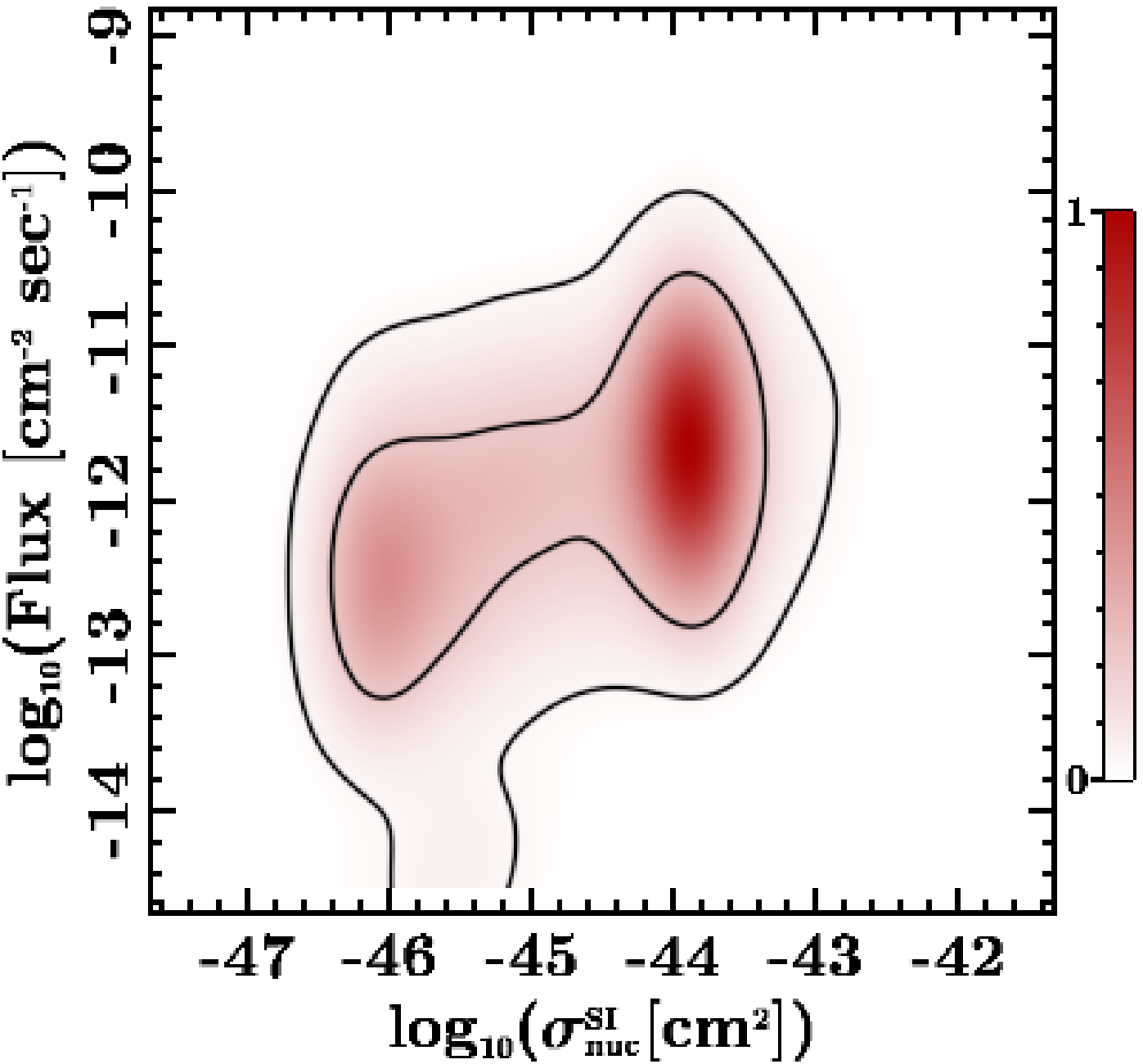}}
\rotatebox{0}{\includegraphics[height=0.45\hsize]{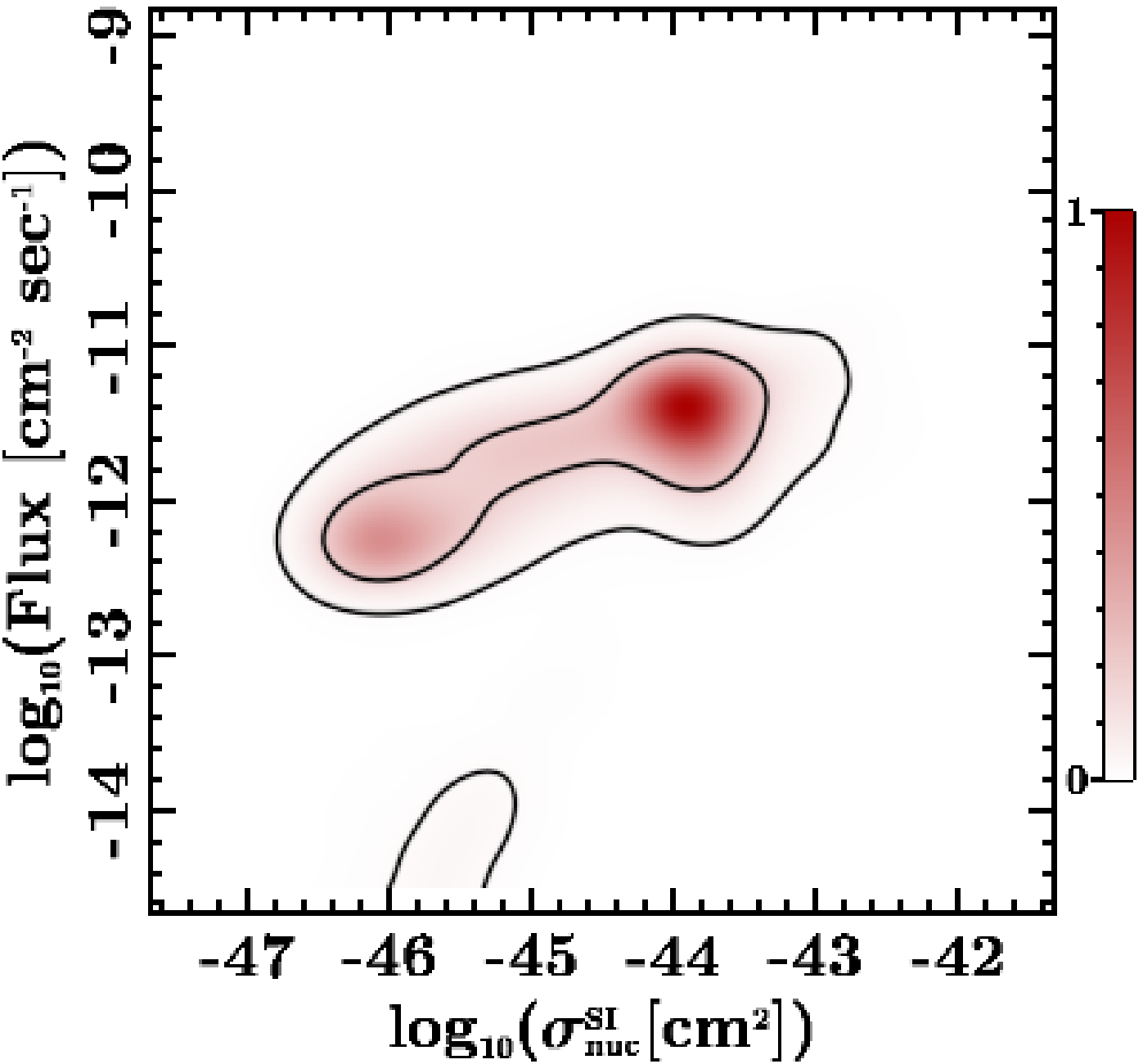}}
\caption{\label{fig:flux_SI}\footnotesize
The $E>1$ GeV gamma-ray flux versus the spin-independent cross section for 
Segue 1 ({\em Left}) and Draco ({\em Right}). Both figures include the
conservative boost model corresponding to
the~\citet{Bullock2001} halo concentration model
(discussed in~\Sref{subsec:boostmodels}) and CDM prior (discussed
in~\Sref{subsec:priors}).
Inner and outer contours represent the 68\% and 95\% c.l. regions
respectively.  
}
\end{center} 
\end{figure}

\begin{figure}
\begin{center}
\rotatebox{270}{\includegraphics[height=0.45\hsize]{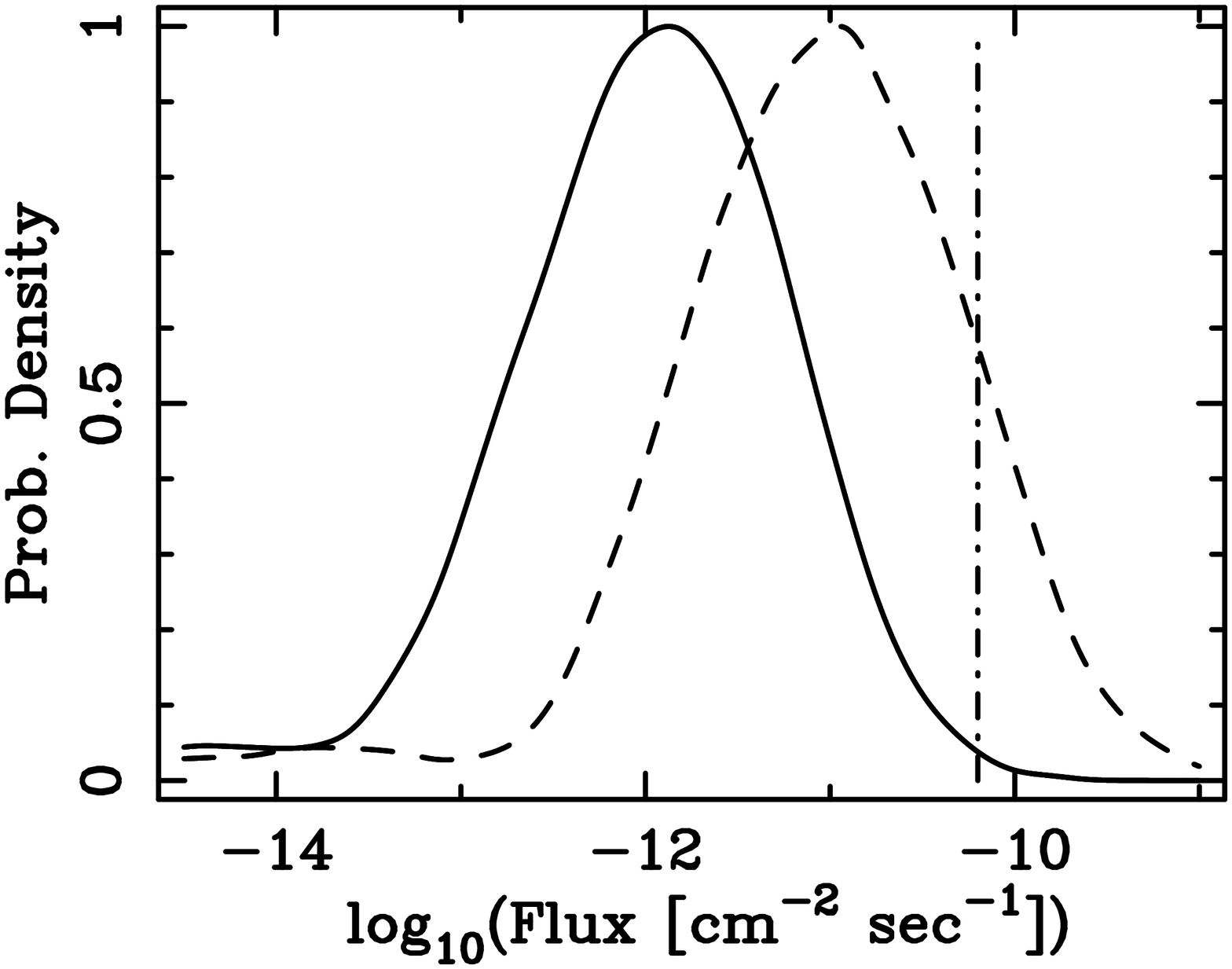}}
\rotatebox{270}{\includegraphics[height=0.45\hsize]{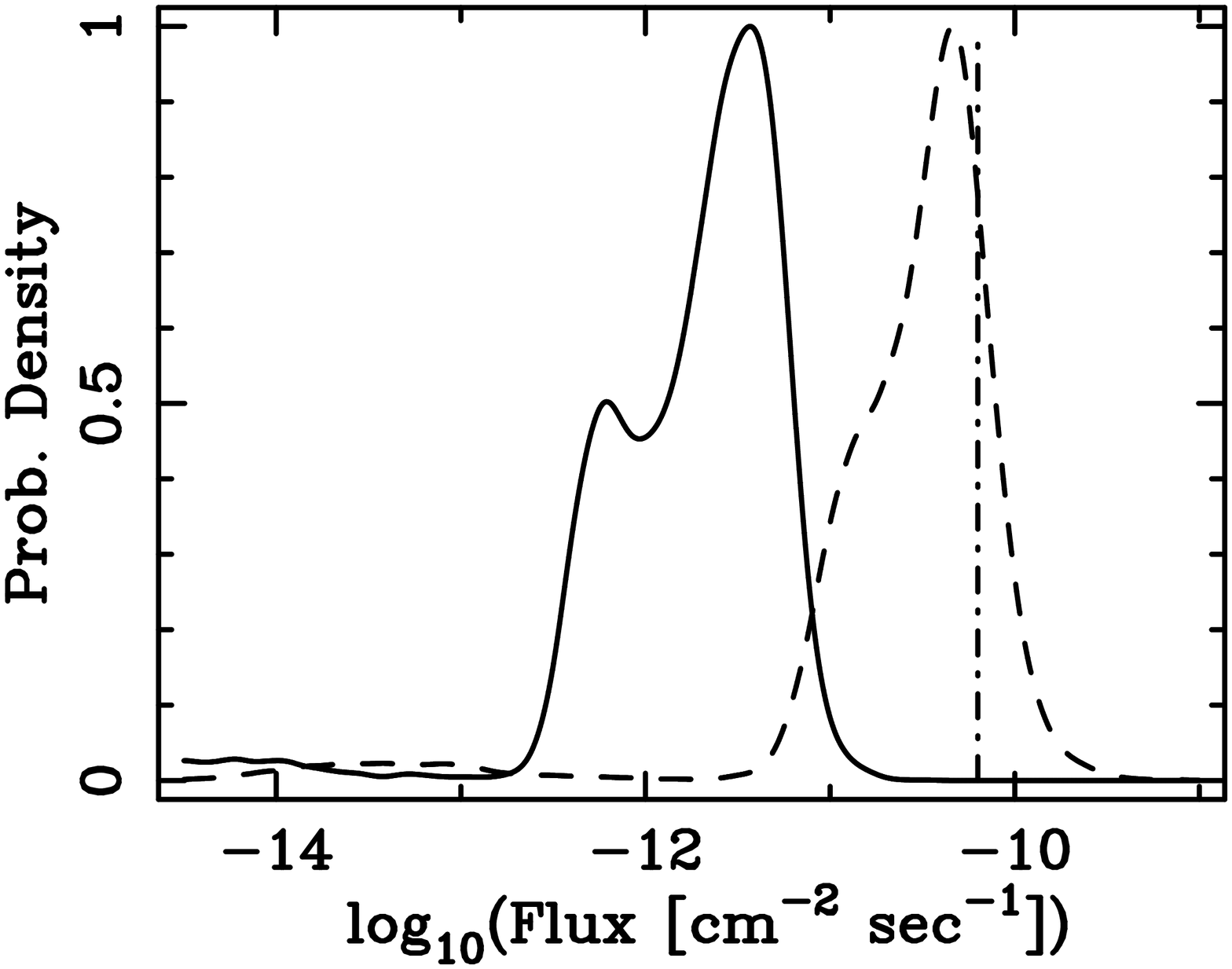}}
\caption{\label{fig:1Dfluxes}\footnotesize
Probability density for the flux of $E>1$ GeV gamma-rays from Segue I
({\em Left}) and Draco ({\em Right}), after marginalizing over all
parameters of the CMSSM and the dark matter halos. The solid and
dashed curves correspond to the boost model with
the~\citet{Bullock2001} and power-law halo concentration models
(discussed in~\Sref{subsec:boostmodels}) respectively.  The dashed-dotted
line denotes the flux above which the signal to noise is about 3, using  
the estimated definition of signal-to-noise is section~\ref{section:results}.
Notice the double peaked feature in the Draco
posterior.  This is due to the fact that the astrophysical component
of the flux is very well constrained by data causing the uncertainty
in the flux to be dominated by the non-Gaussian CMSSM parameter space.
The uncertainty in the flux for Segue 1, though, is  dominated by the
astrophysics and accordingly is very Gaussian in shape. 
}
\end{center} 
\end{figure}

Given the distribution of  $\theta_{\rm max}$ for each galaxy, in~\Fref{fig:flux_mx}
and~\Fref{fig:flux_SI} we present our results for the flux 
of gamma-rays from each galaxy 
with energies greater than $1$ GeV.
In these figures, we show the gamma-ray flux versus two CMSSM parameters, 
the neutralino mass and the spin-independent cross section. For these
figures we assume the ``B01'' boost model, where the minimum value for
the halo mass is self-consistently calculated from the CMSSM parameter 
space. Because it has more line-of-sight velocities, the
flux distribution for Draco is more strongly-constrained. 
As is seen, the predicted fluxes are similar for both galaxies, despite the
fact that Segue 1 is more than a factor three times closer than Draco. The main
reason that these fluxes are similar traces back to the prior used in
~\Eref{eq:CDMprior}; given the velocity dispersion and half-light
radius, Segue 1 prefers to reside in a halo with lower $V_{\rm max}$ 
relative to Draco. The corresponding one-dimensional flux distributions
are show in~\Fref{fig:1Dfluxes}. 


Using the definition of significance, $\sigma$ (defined above), and an assumed 
exposure, we can obtain a rough estimate of detection prospects. As an example, 
we consider the detection prospects at a signal-to-noise greater than 3 
($\sigma = 3$) in five years (same exposure as given previously). 
Comparing to our flux predictions in~\Fref{fig:1Dfluxes}, we see that
with the standard B01 boost model, the parameter space spanned by the 
flux posterior of Draco lies below this estimated flux limit
(vertical dot-dash line) and only a small fraction of the parameter space reaches 
this limit for Segue 1. 
On the other hand, taking the optimistic PL
boost model, in which the mean boost is  $\sim 20$, both Segue 1 and Draco 
exhibit significant regions of their
parameter space that are detectable given our estimated Fermi
sensitivity. Specifically, for the PL boost model, we find
$\sim 20\%$ of the Draco flux parameter space is
detectable with signal-to-noise greater than 3 with five years of Fermi observation
(and similarly $\sim 13\%$ of the Segue 1 parameter space at the same level). 
We note again that the assumed
priors (see below) have a large effect on the predicted flux and our general 
strategy in this work has been to quantify the {\em minimum} expected flux. 
We also note that for our PL model we assumed a spatial distribution for the subhalos that
tracks the underlying host halo mass distribution.  Making this spatial distribution more extended will lower the flux from the central region. Our results for the B01 concentration boost model are broadly consistent 
with the results of \citet{Pieri08}, who also considered annihilation 
flux from Draco but with an energy threshold of 100 MeV. 

For the PL boost model, it would be interesting to also consider the flux 
of gamma-rays from annihilations in the unresolved substructure of 
the Milky Way and ensure that this is not in conflict with the 
measured EGRET background \cite{PieriBertone08}. However, 
such an exercise is hampered by the facts that the Galactic disk could 
significantly change the substructure distribution as well as abundance, 
and there is no systematic way of taking this into account at the 
present time.

\begin{figure}
\begin{center}
\rotatebox{270}{\includegraphics[height=0.45\hsize]{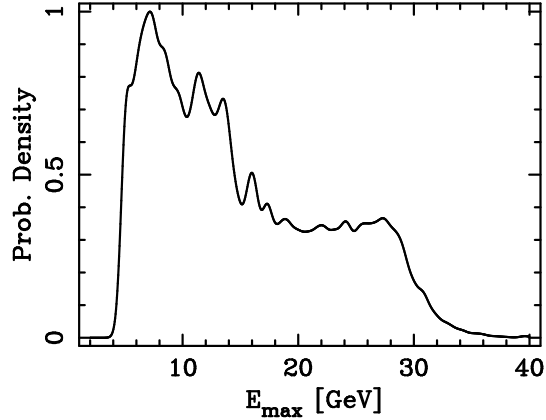}}
\caption{\label{fig:energymax}\footnotesize
The posterior probability density for the peak of the gamma-ray energy spectrum, 
where the energy spectrum is defined in terms of $E^2 dN/dE$. 
}
\end{center} 
\end{figure}

Of course, to extract a dark matter annihilation signal above background,
a detailed comparison between all of the input spectra is required, rather
than simply counting photons above some energy threshold. In order 
to make a detailed comparison between spectra, it will be important
to determine the characterstic energy for photons from dark matter
annihilation. We define the characteristic energy as 
$E_{\rm max}$, the energy at which the quantity $E^2 dN/dE$ peaks. 
For each point in our Markov chain we determine the characteristic energy,  
and the resulting pdf for $E_{\rm max}$
is given in~\Fref{fig:energymax}. 

\begin{figure}
\begin{center}
\rotatebox{270}{\includegraphics[height=0.45\hsize]{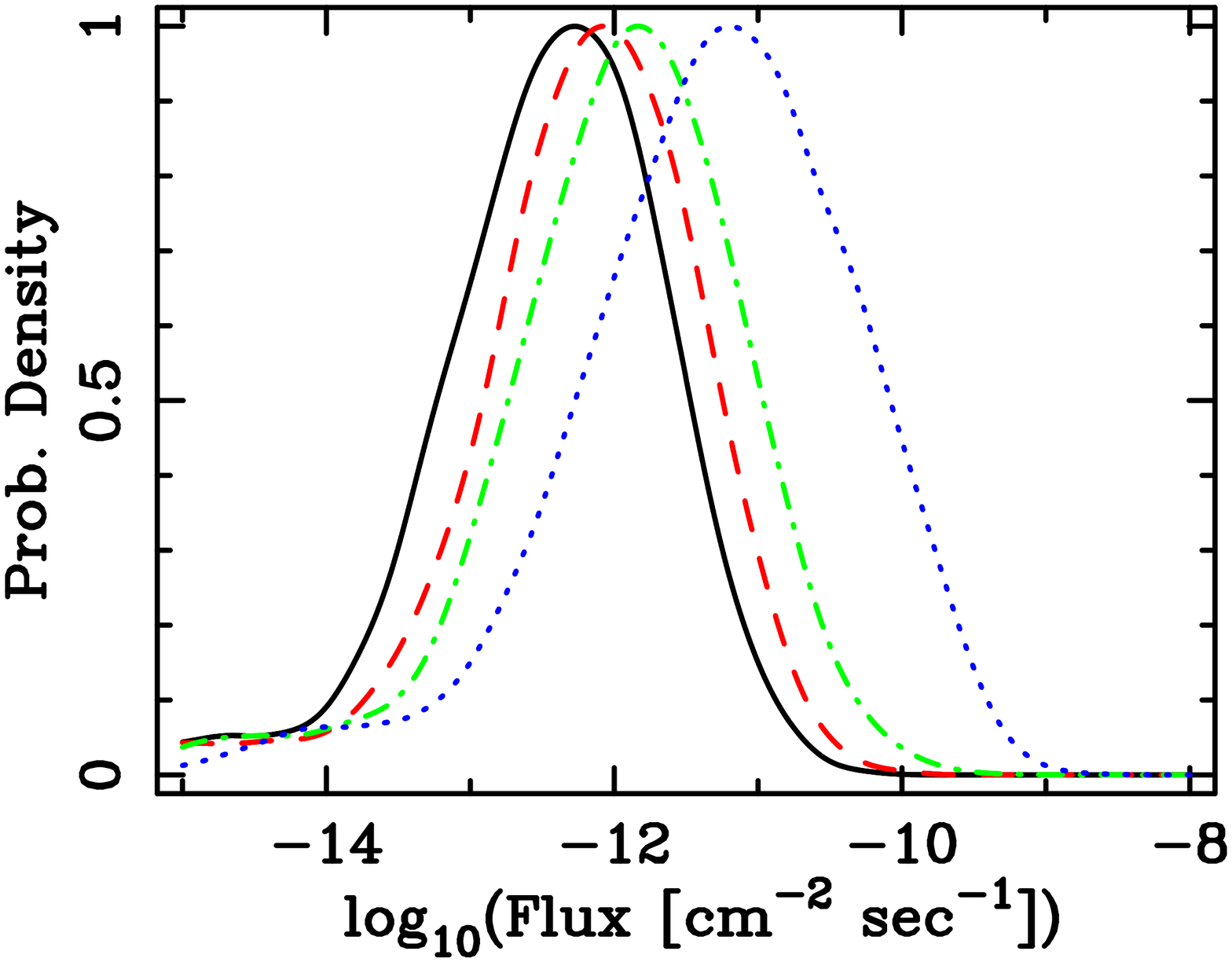}}
\rotatebox{270}{\includegraphics[height=0.45\hsize]{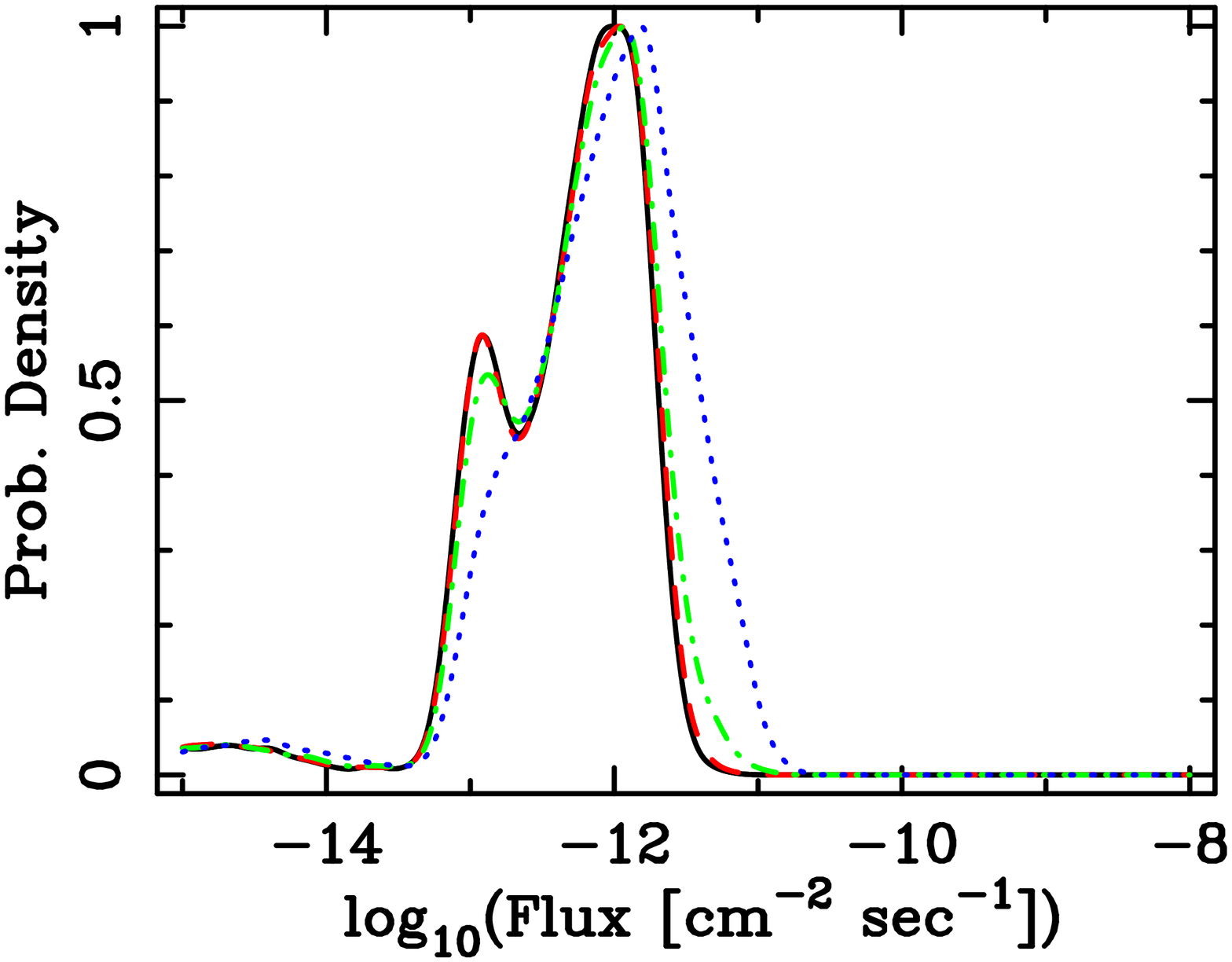}}
\caption{\footnotesize
Effect of the prior on the predicted flux for Segue 1 ({\em Left}) and Draco ({\em Right}).
Again, a uniform prior in $V_{\rm max}^{-3}$ (black, solid)
$V_{\rm max}^{-2}$ (red, dashed), $V_{\rm max}^{-1}$ (green, dot-dashed), 
and $\ln(V_{\rm max})$ (blue, dotted) is assumed.
As in the case of the mass (see~\Fref{fig:prioreffect_seg}), 
flat priors in increasing negative powers of $V_{\rm max}$ causes 
the Segue 1 posterior to be more biased than   
the kinematically well-constrained case of Draco. 
For these figures, we have set the boost factor equal to zero.
\label{fig:fluxprioreffect}
}\end{center} 
\end{figure}

Finally, we re-vist our above discussion of priors in the context of flux
predictions.  We see from \Fref{fig:1Dfluxes} that when the astrophysical data
is well constraining the majority of the flux uncertainty comes from the
particle physics parameter space.  This severely reduces the impact of the astrophysical
priors on the overall uncertainty of the flux.  But for galaxies like Segue 1 with only 24 line-of-sight 
velocities, the likelihood for its parameters is much less constraining and
thus astrophysical priors have a large impact on the flux posterior.
In~\Fref{fig:fluxprioreffect}, we show how the assumed prior affects
the resulting calculation of the flux from Segue 1 and Draco. The curves in~\Fref{fig:fluxprioreffect}
are for the same priors as in~\Fref{fig:prioreffect_seg}. It is thus clear that, given the small
number of stars from Segue 1, the assumed prior has a significant effect
on the flux calculation. The CDM $V_{\rm max}$ prior forces Segue 1 to reside in
halos with smaller $V_{\rm max}$ and this results in smaller predicted fluxes.
We note that the physics of faint-end galaxy formation should change this prior
significantly and would generally push the prior towards favoring larger 
$V_{\rm max}$ values. Thus our chosen priors that have the effect of providing
the minimum expected flux from Segue 1. Future data sets for this
object, and all other ultra-faint satellites, will be crucial for
constraining the dark matter mass and estimating the gamma-ray fluxes.

The priors on the CMSSM parameter space may also play an important role 
in determining the flux expectations from the
dwarfs~\cite{Trotta08}. We have not undertaken a systematic study of
the effect of these priors on the fluxes here. We chose a uniform
prior in all the CMSSM parameters including the masses. 

\section{Discussion and Conclusions \label{section:conclude}}
Constraining the properties of the dark matter particle in indirect detection experiments
will require a firm understanding of the astrophysical uncertainties that contribute to the
flux. Dwarf satellites of the Milky Way are particularly interesting targets in this regards
because they are the most dark matter dominated objects known and they are largely free
from astrophysical uncertainties which result from the presence of baryonic physics. In this
paper, we have taken an step towards quantifying uncertainties in flux predictions from dSphs
by providing a framework within which both particle physics and astrophysics uncertainties
can be included at once.

We have combined our MCMC method for determining the dark matter distributions of  dwarf 
satellites using stellar kinematics with the~\SB~MCMC package which
determines the preferred ranges for parameters of the CMSSM.
We have focused on two specific dSphs, Segue 1 and Draco, as example cases.
Our methods allow us to
provide a broad outline for the prospects of detection of these satellites with
gamma-ray experiments such as Fermi.

The main results of our paper can be summarized as follows: 
\begin{itemize}
\item 
We find that both Draco (at 80 kpc)  and Segue 1 (at 23 kpc) are expected to 
have grossly similar fluxes, though the flux from Segue 1 is subject to larger 
uncertainties because of its relative lack of kinematic data and smaller stellar extent.  
For the most conservative assumptions, Segue 1 prefers to reside in a halo with lower 
$V_{\rm max}$ relative to Draco, and therefore it has a similar overall flux despite its relative 
proximity.  However, for a flat prior in log$(V_{\rm max})$, the flux from
Segue 1 can be much larger than Draco (see Figure~\ref{fig:fluxprioreffect}).  
This result motivates future observations of stellar kinematic data for Segue 1. 
We note that our results for the flux, unless otherwise stated, are based on the
more conservative prior and conservative boost model, and hence they
quantify the minimum expected flux for the assumed CMSSM priors.

\item 
We have provided the first self-consistent calculation for the boost in the
flux signal from halo substructure, taking into account both the CMSSM model  
and the recent results from high resolution numerical simulations.
We show that the dominant uncertainty  in the boost calculation comes from the 
assumed halo concentration versus mass relation for halo substructure on mass 
scales down to the scale of the minimum mass halo.
If we assume a model that links halo concentrations to the power spectrum, then 
we obtain typical boost factors of order unity.  If we instead assume a power
law continuation of the concentration-mass relation down to the minimum mass
halo, the average boost factor increases by an order of magnitude to $\sim 20$ 
(see Figure~\ref{fig:boost}). 
This boost would be reduced if the spatial distribution of the smallest
subhalos is more diffuse that the smooth dark matter halo component, which is
what we have assumed in this paper. We also note that our analytic solution for the boost shows that resolving subhalos, and in some cases sub-subhalos, is sufficient to get an accurate estimate of the boost. These facts motivate future high resolution simulations of halo substructure that can measure the concentrations of small dark matter halos directly, as well as map their spatial distribution within the host. 

\item We have provided a broad outline of the prospects for detection of these satellites
with gamma-ray experiments, focusing specifically on Fermi. We find that, given
the diffuse backgrounds, the most optimal solid angle to view these galaxies is 
$\sim 0.2-0.3^\circ$ (see Figure~\ref{fig:thetamax}). 
Optimistic fluxes for these galaxies are approximately a few times 
$10^{-11}$ cm$^{-2}$ s$^{-1}$. 
For the boost model with the power-law extrapolation of the 
concentration-mass relation and a subhalo spatial distribution that tracks the underlying host mass distribution, 
we estimate a $\sim 20\%$ chance for a $\sim 3\sigma$ dark matter gamma ray signal from Draco after 
5 years of observation. 
This expectation is highlighted in Figure~\ref{fig:1Dfluxes}. We note that 
given the observed uniformity in the central density of the dark matter halos 
of the dwarfs \cite{Strigari:2008ib}, 
it should be possible to stack them and increase the signal-to-noise.

\item 
We have provided an updated calculation for the minimum halo mass in CMSSM. Our results broadly agree with those of~\citet{Profumo:2006bv}, and are typically lower than the masses in~\citet{Bertschinger06} who only considered bino-like neutralinos. We find that the minimum mass
CDM halo lies in the range $10^{-9}-10^{-6}$ M$_\odot$ (see Figure~\ref{fig:boost}).  
The inclusion of this effect results in slightly larger boost estimates. 
\end{itemize}

The methods presented here provide a concrete methodology for addressing the uncertainties 
inherent in dark matter indirect detection from both particle physics and astrophysics perspective.  
The focus on Segue 1 and Draco was meant to illustrate how two vastly different kinematic data 
affect the predictions for the gamma-ray flux from dark matter annihilations. 
It is very interesting to note that Fermi observations of the dwarf satellites of the Milky Way 
could be relevant for constraining the Supersymmetric parameter space.

\ack
We acknowledge support from the NSF for this work through 
grants AST-0607746 and PHY-0555689.
LES is supported by NASA through Hubble Fellowship grant $\#$HF-01225.01
awarded by the Space Telescope Science Institute, which is operated by the Association 
of Universities for Research in Astronomy, Inc., for NASA, under contract NAS 5-26555.

\appendix

\section{MCMC methods and tests}

\subsection{Convergence Test
\label{appendix:convergence}}
\begin{figure}
\rotatebox{270}{\includegraphics[height=0.482\hsize]{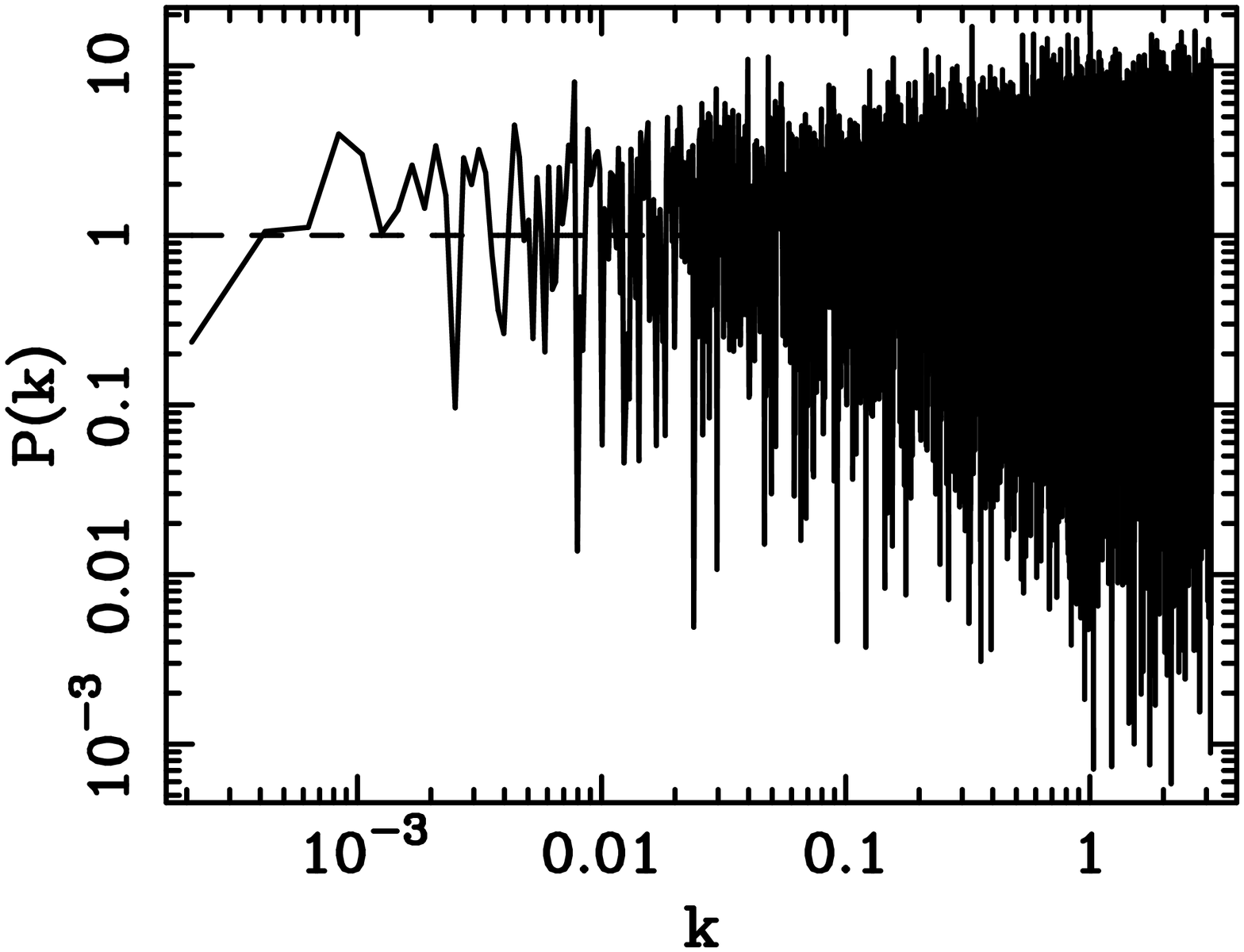}}
\rotatebox{270}{\includegraphics[height=0.482\hsize]{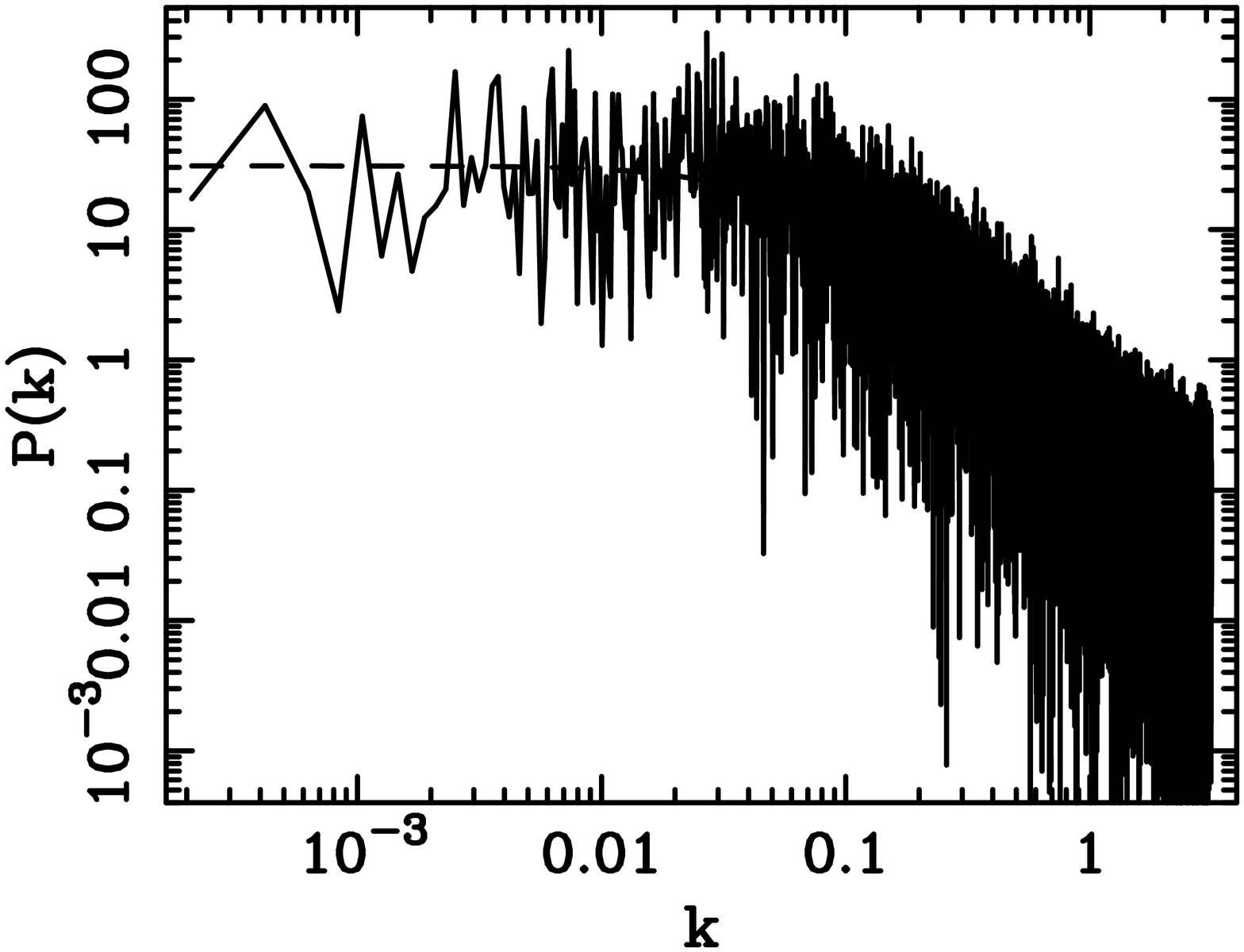}}
\caption{\label{fig:pspec}\footnotesize Power spectra normalized to a
variance of one of a Gaussian random deviate (left), and of our
converged chain for the parameter $r_{\beta}$ (right).  Notice for
small $k$ (large scales) that the MCMC chain becomes flat like that of
the true random deviate.  The dashed lines are the best fit of
~\Eref{eq:pchain}.}
\end{figure}

In this appendix we briefly discuss our criteria for convergence of our
Markov Chains. 

Points close to each other in a 
Monte Carlo Markov Chain are highly correlated and thus do not
represent the distribution being explored.  But, as the chain becomes
larger, points far from each other become
less correlated and eventually become random deviates of the target
distribution.  When this happens, the chain is said to have ``converged." 
This effect can be seen in the
power spectrum of the chain \citep{Dunkley05}.  The power spectrum of
a truly uncorrelated random sample will be flat whereas a random
walk power sectrum has the form $P(k) \propto k^2$ where $k$ is inverse
of the distance between the points.  Thus, for a
converged MCMC chain, the power spectum will be constant at small $k$ 
but will vary as $k^2$ at large $k$ 
(see \Fref{fig:pspec}).  For a converged chain, the power spectrum can
be fitted by the function
\begin{equation}
\label{eq:pchain}
P(k) = P_0 \frac{(k^{\star}/k)^{\alpha}}{1 + (k^{\star}/k)^{\alpha}}.
\end{equation}
Using the best fit values from equation \ref{eq:pchain} the conditions 
for convergence for a chain of size $N$ were set to be
%
\begin{enumerate}
 \item 
\begin{equation}
 j^{\star} \equiv k^{\star}\frac{N}{2 \pi} > 20
\end{equation}
\item
\begin{equation}
 r \equiv \frac{P_0}{N} < 0.01 
\end{equation}
\end{enumerate}
%
where the power spectrum is normalized to a variance of one and
~\Eref{eq:pchain} is fitted over the range $0 < k < 10k^{\star}$ (see
\citet{Dunkley05} for details).  It should be noted however that
this test does not guarantee that the chain has sampled the entire
parameter space, but only that its autocorrelation lengths is sufficiently
short to ensure independence among the samples.

\subsection{Combination Methodology
\label{appendix:combine}}

If two parameter sets are completely independent, then the statistics
of the combined parameter set is independent of whether each parameter
set is derived independently or jointly.  Thus, if a posterior is separable
($\mathcal{P}(\phi, \psi) = \mathcal{P}_{\phi}(\phi)\mathcal{P}_{\psi}(\psi)$),
then the total posterior may be found by the combination of the individual 
distributions.  But care must be taken combining chains with different weights.
If the resulting combined weights differ from the individual distribution weights,
then the resulting combined distribution runs the risk of differing from the original,
pre-combined, distributions when marginalized over.  A solution to this problem is to explicitly repeat
points in each chain (thereby separating the weights out among the repeated points)
in such a way that the two chains sets of weights coinside.  For example, lets consider
the two chains $\phi = \{1, 2, 3, 4\}$ and $\psi = \{a, b, c, d\}$ with weights
$\mathcal{W}_{\phi} = \{4, 3, 1, 1\}$ and $\mathcal{W}_{\psi} = \{1, 2, 3, 3\}$ respectively.  
These chains are equivalent to the lengthier chains of $\phi = \{1, 1, 1, 2, 2, 3, 4\}$ 
and $\psi = \{a, b, c, c, d, d, d\}$ with the common weight set of $\mathcal{W}_{\phi, \psi} = \{1, 2, 1, 2, 1, 1, 1\}$.
Now these two chains can be combined into the merged chain of 
$\{(1, a), (1, b), (1, c), (2, c), (2, d), (3, d), (4, d)\}$ with the above mentioned weight set.
This method will only, at maximum, double the chain length while preserving the individual
parameter distributions. 

\section{Boost Derivation \label{appendix:boost}}

In this appendix, we derive~\Eref{eq:boostans}.  
We start by taking the Laplace transform of~\Eref{eq:dnboost}:
\begin{equation}
sF_n(s) = \frac{A q^{\xi - \alpha}}{s - (\xi-\alpha)} + A q^s F_{n-1}(s)
\end{equation}
where $F_n(s)$ is the Laplace transform of $D_n(b)$.  By recursively substituting $F_n(s)$, we can derive
%
\begin{eqnarray}
\label{eq:fsexpand}
F_n(s) & = & \frac{A q^{\xi - \alpha}}{s(s - (\xi-\alpha))} \nonumber
\left(1 + \frac{A}{s}q^s + \left(\frac{A}{s}q^s\right)^2 + \dots + 
\left(\frac{A}{s}q^s\right)^n F_0(s) \right) \nonumber\\
 & = & \frac{q^{\xi - \alpha - s}}{(s - (\xi-\alpha))}
\sum_{i = 1}^{n} \left(\frac{A}{s}q^s\right)^i.
\end{eqnarray}
%
There are two poles in~\Eref{eq:fsexpand} -- one at $s = 0$ and one at
$s = \xi-\alpha$.  The inverse Laplace transform is the residue at
both these poles.  After a bit of algebra, this becomes
\begin{equation}
\label{eq:boostser}
B_n(M, m_{\rm min}) = \sum_{i = 1}^{n} \left(A q^{\xi - \alpha}\right)^i 
\left[\frac{1}{(i-1)!}\partial_s^{i-1}
\left(\frac{(q^i M/m_{\rm min})^{s- (\xi - \alpha)}}{s - (\xi - \alpha)}\right)
\Bigg|_{s=0} + \frac{1}{(\xi-\alpha)^i}\right]
\end{equation}
where $\partial_s^i$ represents the $i$-th derivative with respect to
$s$.  The first and second terms in~\Eref{eq:boostser} come from the
the residues at $s=0$ and $s=\xi-\alpha$ respectively.
\Eref{eq:boostser} in its current form is very impractical considering
for each term $i-1$ derivatives must be performed.  This motivates
finding a simpler relation for the first term in~\Eref{eq:boostser}.
Letting $c = \xi-\alpha$ and $y = q^i M/m_{\rm min}$,
%
\begin{eqnarray}
\partial_s^{i-1}\left(\frac{y^{s-c}}{s-c}\right)\Bigg|_{s=0} & = & 
\partial_s^{i-1} \left(\int_0^{\ln y} y'^{s-c} d\ln y' + 
\frac{1}{s-c}\right)\Bigg|_{s=0} \nonumber\\
 & = & \int_0^{\ln y} \left(\ln y'\right)^{i-1} y'^{-c} d\ln y' - 
\frac{(i-1)!}{c^i} \nonumber\\
 & = & c^{-i} \gamma(i, c\ln y) - \frac{(i-1)!}{c^i}.
\end{eqnarray}
%
This leads to
\begin{equation}
B_n(M, m_{\rm min}) = \sum_{i=1}^{n} \frac{1}{(i-1)!} 
\left(\frac{A q^{\xi - \alpha}}{\xi - \alpha}\right)^{i}\gamma 
\left(i, (\xi-\alpha)\ln\left(\frac{q^{i} M}{m_{\rm min}}\right)\right).
\end{equation}
The argument in the incomplete gamma function can either be positive or negative and follows the recursion relations
%
\begin{eqnarray}
\gamma(i, c\ln y) & = & (i-1)\gamma(i-1, c\ln y) - \left(c\ln y\right)^{i-1} y^{-c} \\
\gamma(1, c\ln y) & = & 1 - y^{-c}.
\end{eqnarray}
%

\section{Scattering Matrix Calculation \label{appendix:scat}}
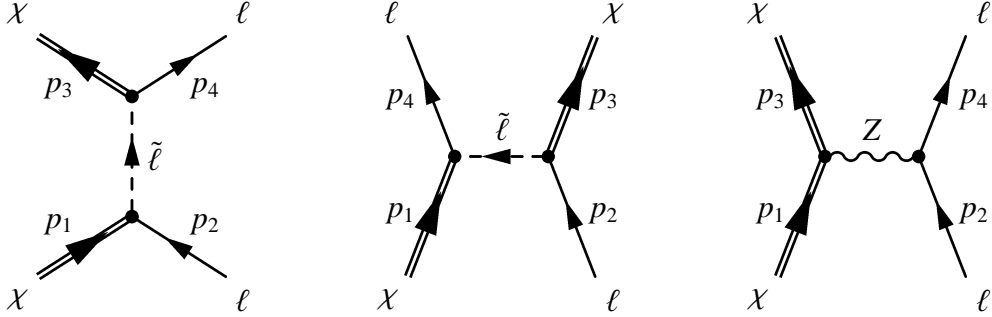
\begin{figure*}
\begin{fmffile}{fmftempl}
\fmfframe(15,15)(5,5){
 \begin{fmfgraph*}(25,40) \label{box}
  \fmftop{sxo,slo} \fmflabel{$\chi$}{sxo} \fmflabel{$\ell$}{slo}
  \fmfbottom{sxi,sli} \fmflabel{$\chi$}{sxi} \fmflabel{$\ell$}{sli}
  \fmf{heavy,label=$p_1$}{sxi,v1}
  \fmf{fermion,label=$p_2$}{sli,v1}
  \fmf{heavy,label=$p_3$,label.side=left}{v2,sxo}
  \fmf{fermion,label=$p_4$,label.side=right}{v2,slo}
  \fmf{scalar,label=$\tilde{\ell}$}{v1,v2}
  \fmfdot{v1,v2}
 \end{fmfgraph*}}
\end{fmffile}
\begin{fmffile}{fmftemp2}
\fmfframe(15,15)(5,5){
 \begin{fmfgraph*}(25,40)
  \fmftop{ulo,uxo} \fmflabel{$\chi$}{uxo} \fmflabel{$\ell$}{ulo}
  \fmfbottom{uxi,uli} \fmflabel{$\chi$}{uxi} \fmflabel{$\ell$}{uli}
  \fmf{heavy,label=$p_1$,label.side=left}{uxi,uv2}
  \fmf{fermion,label=$p_2$,label.side=right}{uli,uv1}
  \fmf{heavy,label=$p_3$,label.side=right}{uv1,uxo}
  \fmf{fermion,label=$p_4$,label.side=left}{uv2,ulo}
  \fmf{scalar,label=$\tilde{\ell}$}{uv1,uv2}
  \fmfdot{uv1,uv2}
 \end{fmfgraph*}}
\end{fmffile}
\begin{fmffile}{fmftemp3}
\fmfframe(15,15)(5,5){
 \begin{fmfgraph*}(25,40)
  \fmftop{uxo,ulo} \fmflabel{$\chi$}{uxo} \fmflabel{$\ell$}{ulo}
  \fmfbottom{uxi,uli} \fmflabel{$\chi$}{uxi} \fmflabel{$\ell$}{uli}
  \fmf{heavy,label=$p_1$,label.side=left}{uxi,uv2}
  \fmf{fermion,label=$p_2$}{uli,uv1}
  \fmf{heavy,label=$p_3$,label.side=left}{uv2,uxo}
  \fmf{fermion,label=$p_4$}{uv1,ulo}
  \fmf{photon,label=$Z$}{uv1,uv2}
  \fmfdot{uv1,uv2}
 \end{fmfgraph*}}
\end{fmffile}
\caption{\label{fig:feyn}\footnotesize
Tree level feynman diagrams for the s, u, and t channels respectively for the process $\chi+ \ell \rightarrow \chi+ \ell$.}
\end{figure*}

In this appendix we derive the cross section for neutralino-lepton
elastic scattering.  We closely follow \citet{Chen01}'s procedure in
finding the neutralino-neutrino scattering matrix element with the
exception that we generalized their result for any lepton.  We start
with the amplitudes for s, u, and t channels represented in
\Fref{fig:feyn} (see \citet{HaberKane85, Gates88, Jungman96} for the
appropriate Feynman rules).  The t channels contributions due to an
Higgs boson exchange become negligible with the assumption of highly
relativistic leptons (e.g. the coupling constants vanish for $m_{\ell}
\rightarrow 0$).  The matrix amplitudes due to each individual diagram
are
%
\begin{eqnarray}
\mathcal{M}_s & = & i\frac{1}{s - m_{\tilde{\ell}_j}^2}\bar{u}_4
\left({X}_{\ell i j n}' P_R + {W}_{\ell i j n}' P_L\right) C
\bar{u}_3^T u^T_1 C^{-1}\left({X}_{\ell i j n}' P_L + {W}_{\ell i j
n}' P_R\right)u_2 \\ \mathcal{M}_u & = & -i\frac{1}{u -
m_{\tilde{\ell}_j}^2}\bar{u}_4 \left({X}_{\ell i j n}' P_R + {W}_{\ell
i j n}' P_L\right) u_1 \bar{u}_3 \left({X}_{\ell i j n}' P_L +
{W}_{\ell i j n}' P_R\right)u_2 \\ \mathcal{M}_t & = &
i\frac{g^2}{2\cos \theta_w}\frac{1}{t-m_Z^2}\bar{u}_4
\gamma^{\mu}\left(c_L P_L + c_R P_R\right) u_2 \bar{u}_3
\gamma_{\mu}\left(O_{nnL}'' P_L + O_{nnR}'' P_R\right)u_1
\end{eqnarray}
%
where
%
\begin{eqnarray}
 P_L & = & \frac{1 - \gamma^5}{2} \\
 P_R & = & \frac{1 + \gamma^5}{2} \\
 c_L & = & \cases{1 & for $\nu_e$, $\nu_{\mu}$, $\nu_{\tau}$ \\ 2 \sin^2 \theta_w-1 & for $e$, ${\mu}$, ${\tau}$ \\} \\
 c_R & = & \cases{0 & for $\nu_e$, $\nu_{\mu}$, $\nu_{\tau}$ \\ 2 \sin^2 \theta_w & for $e$, ${\mu}$, ${\tau}$. \\}
\end{eqnarray}
%
and the coupling constants ${X}_{\ell i j n}'$, ${W}_{\ell i j n}'$,
and $O_{nn\{R,L\}}''$ are defined in reference \citep{Jungman96}.
Here, the indices $\ell$, $i$, $j$, and $n$ represent the family type
(charged lepton or neutrino), the lepton flavor, the slepton flavor,
and the scattering neutralino.  Here, $C = \gamma^2\gamma^0$ is the
charge conjugation matrix.  Using the above amplitudes, the combined
matrix element is
\begin{equation}
\left|\mathcal{M}\right|^2 = \left|\mathcal{M}_s\right|^2+\left|\mathcal{M}_u\right|^2\left|+\mathcal{M}_t\right|^2 + 2\mathcal{R}\left(\mathcal{M}_s\mathcal{M}_u^*+\mathcal{M}_s\mathcal{M}_t^*+\mathcal{M}_t\mathcal{M}_u^*\right)
\end{equation}
where the individual components to the matrix element are
%
\begin{eqnarray}
\left|\mathcal{M}_s\right|^2 & = & 4\left[{\left|\Pi_X^s\right|}^2 +
{\left|\Pi_W^s\right|}^2 + 2{\left|\Pi_{XW}^s\right|}^2\right] (p_1
\cdot p_2)(p_3 \cdot p_4) \\ \left|\mathcal{M}_u\right|^2 & = &
4\left[{\left|\Pi_X^u\right|}^2 + {\left|\Pi_W^u\right|}^2 +
2{\left|\Pi_{XW}^u\right|}^2\right] (p_1 \cdot p_4)(p_2 \cdot p_3) \\
\left|\mathcal{M}_t\right|^2 & = & \frac{4 g^4}{\cos^4 \theta_w (t -
m_Z^2)^2}\Bigg\{\left[\left|O_{nnL}''\right|^2\left| c_{L}\right|^2 +
\left| O_{nnR}''\right|^2\left| c_{R}\right|^2\right](p_1 \cdot
p_2)(p_3 \cdot p_4) \nonumber \\ & & + \left[\left|
O_{nnL}''\right|^2\left| c_{R}\right|^2 + \left|
O_{nnR}''\right|^2\left| c_{L}\right|^2\right](p_1 \cdot p_4)(p_2
\cdot p_3) \nonumber \\ & & \quad -
\mathcal{R}\left(O_{nnL}''{O_{nnR}''}^*\right)\left[\left|
c_R\right|^2 + \left| c_R\right|^2\right]m_{\chi}^2 (p_2 \cdot
p_4)\Bigg\} \\ \mathcal{M}_s\mathcal{M}_u^* & = &
4\Pi_{XW}^s{\Pi_{XW}^u}^*\Big[(p_1\cdot p_2)(p_3\cdot p_4) + (p_1\cdot
p_4)(p_2\cdot p_3) \nonumber \\ & & - (p_1\cdot p_3)(p_2\cdot
p_4)\Big]- 2 m_{\chi}^2(p_2\cdot
p_4)\left[\Pi_{X}^s{\Pi_{X}^u}^*+\Pi_{W}^s{\Pi_{W}^u}^*\right] \\
\mathcal{M}_s\mathcal{M}_t^* & = & -\frac{2g^2}{\cos^2 \theta_w (t -
m_Z^2)}\Bigg\{2\left[{\Pi_X^s} c_L^* {O_{nnL}''}^* + {\Pi_W^s} c_R^*
{O_{nnR}''}^*\right](p_1\cdot p_2)(p_3\cdot p_4) \nonumber \\ & & -
\left[{\Pi_X^s} c_L^* {O_{nnR}''}^* + {\Pi_W^s} c_R^*
{O_{nnL}''}^*\right]m_{\chi}^2(p_2\cdot p_4) \Bigg\} \\
\mathcal{M}_t\mathcal{M}_u^* & = & -\frac{2g^2}{\cos^2 \theta_w (t -
m_Z^2)}\Bigg\{2\left[{\Pi_X^s}^* c_L O_{nnR}'' + {\Pi_W^s}^* c_R
O_{nnL}''\right](p_1\cdot p_4)(p_2\cdot p_3) \nonumber \\ & & -
\left[{\Pi_X^s}^* c_L O_{nnL}'' + {\Pi_W^s}^* c_R
O_{nnR}''\right]m_{\chi}^2(p_2\cdot p_4) \Bigg\}.
\end{eqnarray}
%
Here we used the simplifying notation
%
\begin{eqnarray}
 \Pi_X^c & = & \sum_j\frac{{X}_{\ell i j n}'^2}{c - m_{\tilde{\ell}_j}^2} \\
 \Pi_W^c & = & \sum_j\frac{{W}_{\ell i j n}'^2}{c - m_{\tilde{\ell}_j}^2} \\
 \Pi_{XW}^c & = & \sum_j\frac{{X}_{\ell i j n}'{W}_{\ell i j n}'}{c - m_{\tilde{\ell}_j}}.
\end{eqnarray}
%
In the limit of highly relativistic leptons and by assuming $m_{\chi}
\gg E_{\ell}$, the specific momentum scalar products can be
approximated by
%
\begin{eqnarray}
(p_1 \cdot p_3) \simeq s \simeq u \simeq m_{\chi}^2 \\
(p_1 \cdot p_2) \simeq (p_2 \cdot p_3) \simeq (p_3 \cdot p_4) \simeq (p_1 \cdot p_4) \simeq m_{\chi} E_{\ell} \\
(p_2 \cdot p_4) \simeq -\frac{t}{2} \simeq E_{\ell}^2(1-\cos \theta).
\end{eqnarray}
%
This leads to a matrix element of
\begin{eqnarray}
\left|\mathcal{M}\right|^2 & = & 8 m_{\chi}^2
 E_{\ell}^2\Bigg\{2{\left|\Pi_{XW}^{m_{\chi}^2}\right|}^2 + \left(1 +
 \frac{t}{4
 E_{\ell}^2}\right)\Bigg[{\left|\Pi_X^{m_{\chi}^2}\right|}^2 +
 \left|\Pi_W^{m_{\chi}^2}\right|^2 +
 2{\left|\Pi_{XW}^{m_{\chi}^2}\right|}^2 \nonumber \\ & &
 -\frac{g^2}{\cos^2 \theta_w
 (t-m_Z^2)}\mathcal{R}\left(\left(\Pi_X^{m_{\chi}^2}c_L^*+\Pi_W^{m_{\chi}^2}c_R^*\right)\left({O_{nnL}''}^*+{O_{nnR}''}^*\right)\right)\Bigg]
 \nonumber \\ & & \quad+\frac{g^4\left(\left| c_R\right|^2 + \left|
 c_R\right|^2\right)}{\cos^4 \theta_w (t - m_Z^2)^2}\left(\frac{\left|
 O_{nnL}''\right|^2+\left|
 O_{nnR}''\right|^2}{2}+\frac{\mathcal{R}\left(O_{nnL}''{O_{nnR}''}^*\right)}{4
 E_{\ell}}t\right)\Bigg\}.
\end{eqnarray}
\bibliographystyle{apsrev}

\bibliography{boost}

\end{document}